\documentclass[10pt,journal,twoside,final]{IEEEtran}

\usepackage{diagbox}
\usepackage{graphicx}
\usepackage[caption=false,font=footnotesize]{subfig}

\usepackage{amssymb}
\usepackage{amsmath}
\usepackage{url}

\usepackage{enumitem}

\usepackage{multirow}
\usepackage{multicol}
\usepackage{bigstrut}
\usepackage{booktabs}

\DeclareMathOperator*{\argmin}{argmin}

\newcommand{\etal}{\textit{et al. }}

\AtBeginDocument{%
  \providecommand\BibTeX{{%
    \normalfont B\kern-0.5em{\scshape i\kern-0.25em b}\kern-0.8em\TeX}}}

\begin{document}
%
\title{Secrets in Radio Waves: Towards Practical and Protocol-Agnostic PHY Information Hiding}

\author{
    Guanxiong~Shen,
    Hailang~Jia,
    Junqing~Zhang,~\IEEEmembership{Senior~Member,~IEEE},
    Linning~Peng,~\IEEEmembership{Member,~IEEE},
    Liquan~Chen,~\IEEEmembership{Member,~IEEE},
    Aiqun~Hu,~\IEEEmembership{Member,~IEEE},
    Jun~Luo,~\IEEEmembership{Fellow,~IEEE}
	
	\thanks{This is the accepted version of a paper to appear in \textit{IEEE Transactions on Mobile Computing}. \copyright~2026 IEEE. Personal use of this material is permitted. Permission from IEEE must be obtained for all other uses, in any current or future media, including reprinting/republishing this material for advertising or promotional purposes, creating new collective works, for resale or redistribution to servers or lists, or reuse of any copyrighted component of this work in other works. \textit{(Corresponding author: Aiqun Hu.)}}

	
	\thanks{G.~Shen, H.~Jia, L.~Peng, and L.~Chen are with the School of Cyber Science and Engineering, Southeast University, Nanjing, 210096, China (e-mail: gxshen@seu.edu.cn; 220245603@seu.edu.cn; pengln@seu.edu.cn; lqchen@seu.edu.cn).}

	\thanks{J.~Zhang is with the School of Computer Science and Informatics, University of Liverpool, Liverpool, L69 3DR, U.K. (e-mail: Junqing.Zhang@liverpool.ac.uk).}

	\thanks{A.~Hu is with the School of Information Science and Engineering, Southeast University, Nanjing, 210096, China (e-mail: aqhu@seu.edu.cn).}

	\thanks{J.~Luo is with the School of Computer Science and Engineering, Nanyang Technological University, 639798, Singapore (e-mail: junluo@ntu.edu.sg).}
}

\maketitle

\begin{abstract}

Physical layer (PHY) information hiding supports critical applications, such as digital fingerprinting for transmitter identification and undetectable side channels for covert communication, and has attracted considerable attention from the research community. 
One category of prior studies focuses on theoretical analysis, proposing techniques such as artificial noise or reconfigurable intelligent surfaces to enable undetectable covert transmission. However, hardware prototypes are rarely presented due to their algorithmic complexity or hard-to-satisfy assumptions.
Another category of studies focuses on system-level solutions, achieving PHY information hiding by customizing existing modulation schemes. However, these methods are typically designed for specific wireless protocols, limiting their generalizability.
In this work, we introduce a new PHY information hiding paradigm that differs fundamentally from the previous two categories of approaches. Inspired by recent advancements in other domains such as image information hiding, we migrate encoder-decoder neural networks to the PHY information hiding field, embedding secrets by introducing imperceptible distortions within the preamble waveform. Sim-to-real fine-tuning is additionally proposed to tackle unique challenges, e.g., fading and hardware imperfections. The designed methodology is practical and protocol-agnostic. We provide case hardware prototypes of two commercially popular wireless technologies, i.e., LoRa and Bluetooth Low Energy (BLE), using commodity software-defined radio (SDR) transceivers, demonstrating excellent feasibility and generalizability.

\end{abstract}





\section{Introduction}\label{sec:introduction}

Information hiding is a security technique that conceals secrets within other ordinary media such as images and texts~\cite{petitcolas2002information}. Its practical applications include digital watermarking for intellectual property protection~\cite{luo2020distortion}, fingerprint embedding for authentication~\cite{paul2015wireless, ito2023cancelable}, and steganography for covert communications~\cite{petitcolas2002information, katzenbeisser2016information}.
While conventional information hiding techniques primarily focus on hiding secrets within images, videos, texts, or audio, their application in wireless systems has recently gained increasing attention, particularly at the physical layer (PHY).
Concealing secrets within PHY radio waveforms offers distinct advantages over hiding information within payload contents. Specifically, given that most off-the-shelf wireless receivers lack access to raw radio waveforms, potential eavesdroppers have to deploy specialized hardware and steganalysis algorithms to detect covert channels, which significantly raises the cost and complexity of launching successful attacks. This characteristic additionally provides inherent stealthiness and imperceptibility.
PHY information hiding enables critical applications such as digital fingerprint embedding for transmitter authentication~\cite{paul2015wireless} and establishing side channels for covert communication~\cite{classen2015practical}. Consequently, advancing PHY information hiding techniques is of significant importance.

The wireless community has shown considerable interest in developing PHY information hiding and covert communication schemes~\cite{chen2023covert}. Existing studies are roughly divided into two categories, i.e., theory-oriented and system-oriented. The former category of studies focuses on theoretical analysis, utilizing multiple antennas~\cite{shmuel2021multi,du2022performance}, reconfigurable intelligent surface (RIS)~\cite{wang2021intelligent,lu2020intelligent}, or mmWave~\cite{zhang2021joint, wang2021covert} to achieve directional transmission, thus minimizing signal leakage to the eavesdropper. However, their validation is typically limited to numerical simulations due to algorithmic complexity or hard-to-satisfy assumptions. In contrast to these theoretical works, system-oriented studies aim to present practical hardware prototypes based on existing wireless protocols~\cite{hou2022cloaklora, liu2023lophy, schulz2018shadow, huang2019reliable, bonati2021stealte}. For instance, the authors in~\cite{hou2022cloaklora} design a wireless steganography technique for the LoRa protocol, where covert messages are embedded by modulating the amplitudes of chirp signals. Similarly, Huang~\etal propose a radio frequency (RF) watermarking scheme for narrow-band Internet of Things (NB-IoT) systems~\cite{huang2019reliable}, which embeds secrets by shifting the I/Q constellations. However, although these approaches are experimentally validated as effective, they are often designed for a specific wireless protocol or modulation scheme, resulting in insufficient generalizability.
The development of a protocol-agnostic PHY information hiding framework requires further exploration.

The paradigm of information hiding for images~\cite{zhu2018hidden, luo2020distortion, lu2021large} and text~\cite{abdelnabi2021adversarial, yang2018rnn,kaptchuk2021meteor} has been fundamentally changed by deep learning in recent years, whereas its application in PHY information hiding remains unexplored due to unique challenges. 
The deep learning-driven information hiding methods typically employ a pair of neural networks, i.e., an encoder-decoder framework, to facilitate secret embedding and extraction~\cite{wang2023data}. Specifically, Alice utilizes a neural network to embed the secret into a carrier medium, while Bob employs a paired network to extract the hidden information. This approach has proven highly effective in domains such as images and texts.
However, to the authors' best knowledge, the encoder-decoder framework has never been explored in the literature on PHY information hiding. This methodological gap is due to the unique challenges of hiding secrets within radio waves. Unlike images or text, which can be conveyed without significant distortion, radio waves propagate through the air and are inevitably affected by surrounding environments, i.e., fading and hardware imperfections. It is challenging to ensure the survival of secrets after experiencing environmental fading and hardware distortions. Moreover, it is equally critical to ensure the primary communication link remains unaffected by the embedded secrets.

To tackle the above challenges, we propose to embed the secrets within the preamble waveforms.
Specifically, we design a pair of encoder-decoder neural networks that can process arbitrary lengths of I/Q samples for secret embedding and extraction.
A channel simulator is integrated during the training process, and a sim-to-real fine-tuning scheme is proposed to mitigate fading and hardware distortions. Our main contributions are highlighted as follows:
\begin{itemize}
    \item We propose a PHY information hiding methodology that employs a neural network to embed secret information as subtle distortions within the preamble waveform, fundamentally differing from prior studies. This methodology enables the design of effective PHY-layer information hiding algorithms through simulation-based learning and is applicable to any wireless communication protocol.
    
    \item A pair of neural networks is designed for secret embedding and extraction, capable of processing arbitrary lengths of I/Q samples. A simulation-driven joint training scheme with customized loss functions is introduced, complemented by a sim-to-real fine-tuning technique to mitigate performance degradation when deployed on hardware platforms.
    
    \item The proposed scheme is demonstrated on commercially popular narrowband technologies, i.e., LoRa and BLE. Through extensive simulations, we validate the feasibility of the proposed approach. The results demonstrate excellent PHY information hiding performance when the signal-to-noise ratio (SNR) exceeds 10~dB. While the present work focuses on narrowband validation, the framework is extensible to wideband protocols such as Wi-Fi and 5G by adapting the neural network architecture to the target preamble structure.
    \item We implement functional PHY information hiding prototypes using off-the-shelf software-defined radios (SDRs) for both LoRa and BLE protocols. Experimental results confirm that secret messages can be reliably decoded at the SDR receiver. Furthermore, we validate the scheme's transparency by utilizing a commodity LoRa receiver and a smartphone to receive the SDR-transmitted LoRa and BLE packets, respectively, demonstrating that the primary communication links remain unaffected by the embedded secrets.
    
\end{itemize}
The code, dataset, simulation tools, and reproducible prototypes are publicly available at \url{https://github.com/gxhen/RF_steganography}.

The rest of the paper is organized as follows: Section~\ref{sec:background} presents the background, motivation, and challenges of this work. Section~\ref{sec:system_overview} provides the system overview, and Section~\ref{sec:rf_steganography_system} details the key components in a PHY information hiding system. 
The neural network joint-training and sim-to-real fine-tuning schemes are then introduced in Section~\ref{sec:joint_training}. Section~\ref{sec:real_world_evaluation} presents experimental evaluations of LoRa and BLE PHY information hiding prototypes, respectively. Section~\ref{sec:related_work} reviews related work, discusses applications of PHY information hiding, and highlights our technical advantages. Section~\ref{sec:conclusion} finally concludes the paper.

\section{Background and Motivation}\label{sec:background}

This section first provides background knowledge on information hiding, as well as the LoRa PHY and BLE PHY. After that, we discuss the motivation and key challenges behind our research. Finally, we propose an intuitive solution for designing a practical and protocol-agnostic PHY information hiding system.

\subsection{Background Knowledge}\label{sec:info_hiding_basics}

\begin{figure}[!t]
    \centering
    \includegraphics[width = 3.2in]{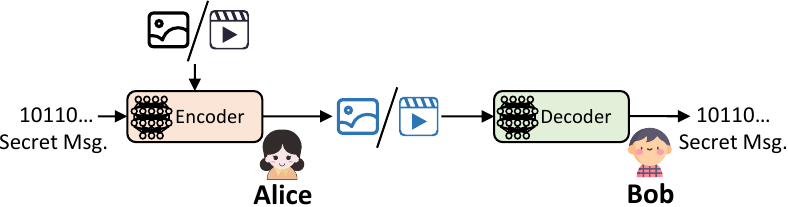}
    \caption{Encoder-decoder information hiding framework. Secret embedding and extraction are achieved using a pair of neural networks.}
    \label{fig:system_model}
\end{figure}

\textit{\textbf{Information hiding}} is an ancient technique that conceals secret messages within other ordinary media, e.g., images~\cite{zhu2018hidden, luo2020distortion, tancik2020stegastamp, lu2021large, bai2024information, tang2019cnn, zhang2020udh} and texts~\cite{yang2018rnn,kaptchuk2021meteor}. The recent advances in deep learning have fundamentally changed the paradigm of information hiding~\cite{wu2020audio, baluja2017hiding, baluja2019hiding, pan2021seek}. As illustrated in Fig.~\ref{fig:system_model}, the deep learning-driven information hiding system employs a pair of neural networks for secret embedding and extraction. Specifically, Alice utilizes a neural network encoder to embed the secrets within a covert medium, e.g., image or video. Subsequently, Bob employs another paired decoder to extract the hidden secrets. This encoder-decoder framework was first proposed in~\cite{zhu2018hidden} to hide secrets within images. It was subsequently rapidly developed, and numerous variants have emerged for image, video, and text information hiding, becoming the \textit{de facto} mainstream in these fields~\cite{zhang2019steganogan, yang2019embedding, wang2023data, yu2024cross, jois2024pulsar}.  However, its effective application in the wireless PHY remains unexplored.

\begin{figure}[!t]
	\centering
	\subfloat[]{\includegraphics[width=1.12in]{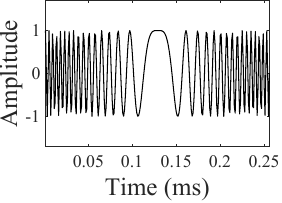}}
	\subfloat[]{\includegraphics[width=1.12in]{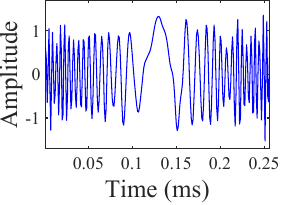}}
	\subfloat[]{\includegraphics[width=1.12in]{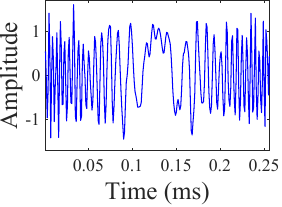}}

	\subfloat[]{\includegraphics[width=1.12in]{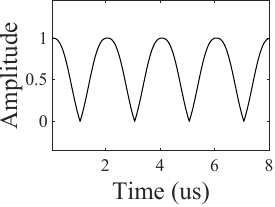}}
	\subfloat[]{\includegraphics[width=1.12in]{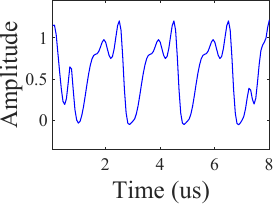}}
    \subfloat[]{\includegraphics[width=1.12in]{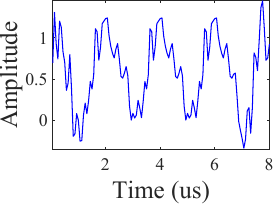}}
	\caption{Visualization of preamble waveform (I-branch). (a) Standard LoRa preamble. (b) LoRa preamble with a 60-bit secret embedded. (c) LoRa preamble with a 180-bit secret embedded. (d) Standard BLE preamble. (e) BLE preamble with a 30-bit secret embedded. (f) BLE preamble with a 120-bit secret embedded.}
	\label{fig:preamble_waveform}
\end{figure}

\textit{\textbf{LoRa}} is a low-power wide area network (LPWAN) technology developed from the chirp spreading spectrum (CSS) modulation technique. It is patented by Semtech and has been employed in numerous IoT applications where long-range communication is required.
A LoRa preamble, i.e., an unmodulated upchirp, is illustrated in Fig.~\ref{fig:preamble_waveform}(a). Typically, a LoRa packet starts with eight repeating preambles, which play a crucial role in synchronization and packet detection.

\textit{\textbf{BLE}} is a short-range wireless personal area network (WPAN) technology that utilizes the Gaussian frequency shift keying (GFSK) modulation scheme at the PHY layer. BLE specifies several PHY modes to support varying data rate requirements. In the basic LE1M mode, the wireless packet includes a preamble consisting of a fixed `10101010' bit sequence for packet detection and synchronization. A standard BLE preamble waveform is shown in Fig.~\ref{fig:preamble_waveform}(d).

\subsection{Motivation, Challenges and Solutions}\label{sec:motivation_challenges}

This study focuses on embedding secrets directly into radio waveforms, rather than relying on traditional cover media such as images or videos, and designs a practical and protocol-agnostic PHY information hiding framework.
The motivation is not to replace higher-layer content-based steganography.
Instead, PHY information hiding attaches secret information directly to the emitted RF waveform while keeping the packet payload and standard receiver operations unchanged.
Since the proposed method embeds information into the radio waveform rather than the packet bits, it remains independent of the carried data type and provides a protocol-agnostic hiding capability.

The applications of the proposed framework, including RF fingerprint embedding and a covert side channel, are discussed in Section~\ref{sec:potential_applications}.

The preamble is a natural carrier: it is present in every packet and has a fixed pattern.
Embedding secrets into the preamble leaves the payload waveform unchanged, so that standard receivers can demodulate the primary communication link without modification.
This paper focuses on preamble-only embedding.
Payload-waveform embedding would require a dedicated design, because both the payload length and the payload content vary from packet to packet.

It is important to emphasize that directly migrating algorithms from studies targeting images~\cite{zhu2018hidden} and text~\cite{kaptchuk2021meteor} may not be effective.
In contrast to images and videos that can be conveyed without distortion, radio waves propagate through the air. The surrounding environments and transceiver hardware imperfections can result in severe fading and distortions within the received signals. It is challenging to ensure the survival of embedded secrets after the radio signal is distorted. 
Another critical consideration is to ensure the primary communication link remains unaffected by the created covert side channel, which further complicates the system design.
This work aims to answer four questions: 
\begin{itemize}
    \item[i)] How to design a practical and protocol-agnostic PHY information hiding paradigm?
    \item[ii)] How to effectively migrate encoder-decoder neural networks into PHY information hiding?
    \item[iii)] How to ensure the survival of secrets after the radio undergoes fading and hardware distortions?
    \item[iv)] How to ensure the primary communication link remains unaffected by the embedded secrets?
\end{itemize}

To this end, a pair of neural networks is employed, one for embedding the secrets into the preamble and another for extracting them at the receiving end. The distorted LoRa and BLE preambles, with secrets embedded, are visualized in Fig.~\ref{fig:preamble_waveform}. As observed, the secret embedder introduces slight distortions to the preamble waveform. These distortions become more perceptible as the number of embedded bits increases. However, the synchronization algorithms remain valid. This is because the distorted waveforms are highly similar to the standard preamble, and the repetition property, i.e., a key feature relied upon by synchronization algorithms, is preserved. Consequently, both cross-correlation and autocorrelation-based synchronization algorithms are expected to remain functional on the receiver side, ensuring successful synchronization despite the embedded secrets. This is validated through both simulation and experiments in Sections~\ref{sec:simulation_evaluation} and~\ref{sec:real_world_evaluation}.

\section{System Overview}\label{sec:system_overview}

\begin{figure*}[!t]
    \centering
    \includegraphics[width = 6.6in]{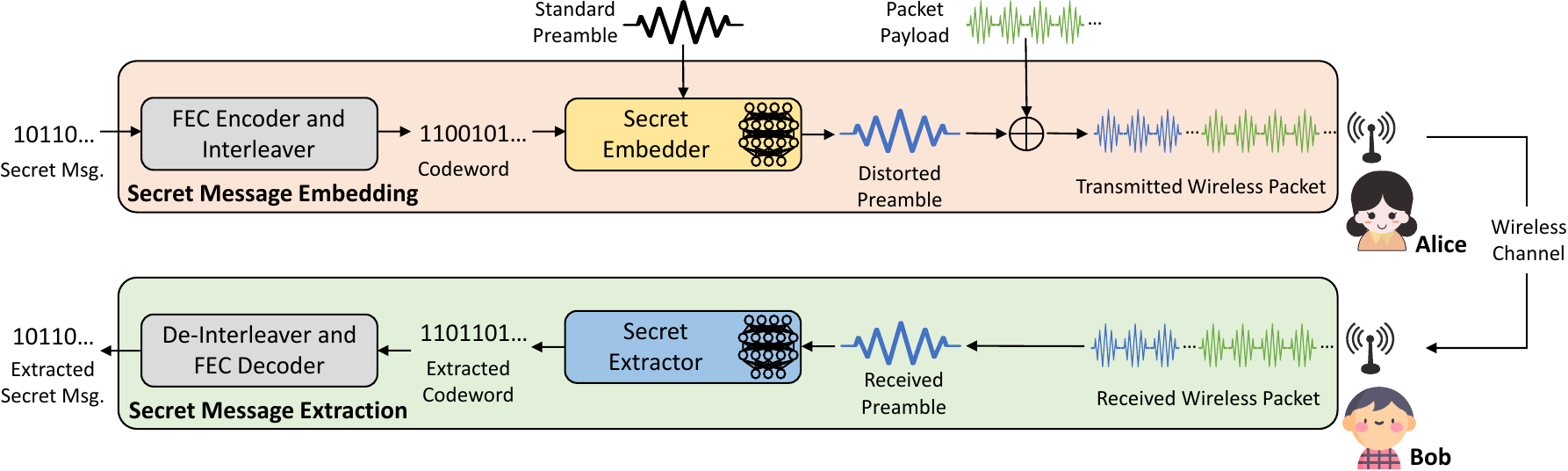}
    \caption{Overview of the proposed PHY information hiding system. }
    \label{fig:system_overview}
\end{figure*}

As illustrated in Fig.~\ref{fig:system_overview}, the proposed PHY information hiding system comprises two parties, namely Alice and Bob. They have already established a primary communication link, such as LoRa, BLE, and ZigBee. Alice aims to hide additional covert messages within the transmitted wireless packets. In more detail, the transmitter, Alice, conceals secret messages within the radio waveforms by introducing subtle and imperceptible distortion using a pre-trained neural network called the secret embedder. 
This secret embedding process only applies to the packet preamble waveform. 
After Bob receives the radio signal, another neural network called the secret extractor is used to extract the secret messages. Forward error correction (FEC) is utilized to correct transmission errors and enhance information hiding performance. The specific operations on Alice and Bob's sides are introduced below.

\textbf{Secret Message Embedding at Alice Side:}
Alice first performs FEC encoding and interleaving to transform the $M$-bit secret message into an $N$-bit codeword. Subsequently, the codeword is embedded into a standard preamble using a pre-trained neural network secret embedder, which introduces subtle and stealthy waveform-level distortions through channel-wise concatenation. The distorted preamble is then concatenated with the payload to generate a complete wireless packet. This is then up-converted from the baseband to the RF band and emitted into the air. The process will be introduced in detail in Section~\ref{sec:Alice}.

\textbf{Secret Message Extraction at Bob Side:}
Bob captures the RF signal, down-converts it to baseband, and digitalizes it with an ADC. Then another pre-trained neural network secret extractor is employed to extract the codeword hidden within the packet preamble field. The extracted codeword is then fed into a de-interleaver and an FEC decoder for error correction, thereby recovering the $M$-bit secret message sent by Alice. The process will be presented in detail in Section~\ref{sec:Bob}.

It is worth noting that the two neural networks, i.e., secret embedder and extractor, are pre-trained jointly in a controlled environment and are then secretly shared between Alice and Bob. The model weights should be protected and kept secret. The details about simulation-driven training and sim-to-real fine-tuning are given in Section~\ref{sec:joint_training} and Section~\ref{sec:fine_tune}, respectively.

PHY information hiding offers a method to establish an undetectable covert communication link between wireless transceivers. The reason for this undetectability is twofold. Firstly, most commodity wireless receivers lack the capability to access physical layer waveforms, i.e., I/Q samples. Those attempting to eavesdrop must possess a bespoke wireless receiver, for example, an SDR, to detect the covert channel. This considerably increases the cost required for eavesdropping. Secondly, the intentionally introduced distortion is subtle and almost imperceptible. Although it is difficult to absolutely guarantee the distortion is undetectable under the observation of high-end equipment, the development of dedicated steganalysis algorithms significantly increases the complexity of the attack. In addition to the undetectability, the eavesdropper has to steal the weights of the secret extractor to parse the embedded covert message, which further increases the cost of the attack.

\section{PHY Information Hiding System}\label{sec:rf_steganography_system}

\subsection{Secret Message Embedding at Alice Side}\label{sec:Alice}

The goal of Alice is to embed a secret message $\mathbf{x}$ into the wireless packet by introducing waveform-level distortion, while not affecting the quality of the primary communication link. This is achieved by the modules detailed below.

\subsubsection{Forward Error Correction Encoder and Interleaver}
Given an $M$-bit secret message $\mathbf{x}$, Alice first performs FEC encoding to introduce redundancy, given as
\begin{equation}
    \mathbf{c} = \mathcal{F}_{\mathrm{enc}}(\mathbf{x}),
\end{equation}
where $\mathcal{F}_{\mathrm{enc}}(\cdot)$ denotes the FEC encoding process and $\mathbf{c}$ is the encoded $N$-bit codeword. Subsequently, Alice utilizes a random interleaver to rearrange the elements in $\mathbf{c}$, thereby mitigating the impact of burst errors that may occur during transmission.

In the LoRa and BLE case studies, the (15, 5) Bose-Chaudhuri-Hocquenghem (BCH) code is employed. This widely used cyclic error correction code is capable of correcting up to three bit errors per codeword. Other techniques, such as low-density parity-check (LDPC) codes, are also feasible.

\subsubsection{Secret Embedder}

\begin{figure}[!t]
    \centering
    \includegraphics[width = 3.3in]{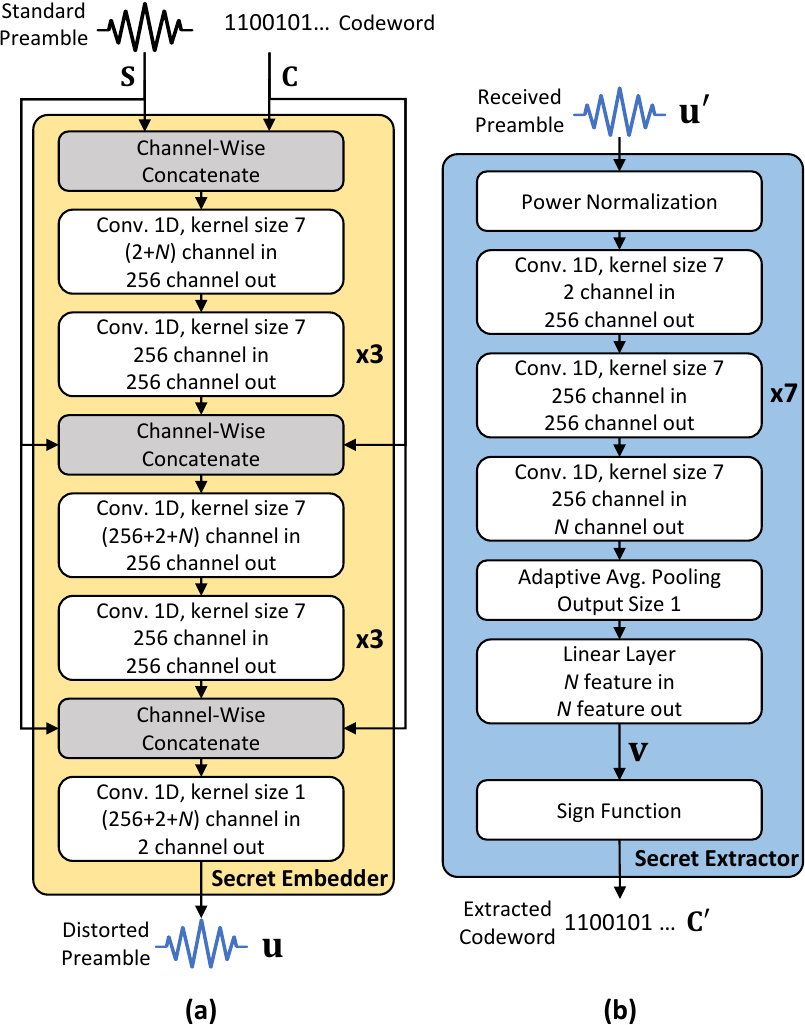}
    \caption{Architecture of the neural network-based secret embedder and extractor. (a) Secret embedder. (b) Secret extractor. }
    \label{fig:nn_architecture}
\end{figure}

After converting the $M$-bit secret message $\mathbf{x}$ to an $N$-bit codeword $\mathbf{c}$, Alice utilizes a pre-trained secret embedder to embed it into the preamble $\mathbf{s}$, which is expressed as 
\begin{equation}
    \mathbf{u} = \mathcal{G}_{\mathrm{emb}}(\mathbf{s}, \mathbf{c} ; \theta_{\mathrm{emb}}),
\end{equation}
where $\mathcal{G}_{\mathrm{emb}}(\cdot,\cdot;\theta_{\mathrm{emb}})$ denotes the neural network-based embedding process and $\theta_{\mathrm{emb}}$ is the parameter set. It accepts two inputs, i.e., a standard preamble $\mathbf{s}$ and a codeword $\mathbf{c}$, and outputs a distorted preamble $\mathbf{u}$. 

The architecture of the designed secret embedder is illustrated in Fig.~\ref{fig:nn_architecture}(a), which involves nine 1D convolutional layers, i.e., a fully convolutional design. The neural network inputs consist of a standard preamble $\mathbf{s}$ and a codeword $\mathbf{c}$. We first perform channel-wise concatenation, combining $\mathbf{s}$ and $\mathbf{c}$ into a single tensor of size ($L$, 2+$N$). This channel-wise concatenation approach is inspired by the image steganography scheme proposed in~\cite{zhu2018hidden} and is further optimized for RF signals. Specifically, the preamble $\mathbf{s}$, which is a complex vector, is split into real and imaginary parts and serves as two separate channels of the tensor. The codeword $\mathbf{c}$, which is an $N$-bit binary vector, is duplicated $L$ times along the time-dimension and serves as $N$ channels of the tensor. The combined tensor is then processed by four 1D convolutional layers, with the channel-wise concatenation applied to the output of the fourth layer again to enhance the embedding process. Note that this channel-wise concatenation has three inputs: the standard preamble $\mathbf{s}$, codeword $\mathbf{c}$, and the output tensor of the fourth convolutional layer, producing a tensor of size ($L$, 256 + 2 + $N$). Subsequently, four more 1D convolutional layers and channel-wise concatenation are applied to further enhance the message embedding performance. Finally, a 1D convolutional layer with a kernel size of one is utilized, which produces an output tensor of size ($L$, 2). It represents the real and imaginary parts of the distorted preamble $\mathbf{u}$, respectively. The outputs of all convolutional layers are activated by GELU functions. 

Thanks to the fully convolutional design, the secret embedder can hide secrets within preambles of arbitrary length and is applicable to both LoRa and BLE preamble waveforms. However, it should be noted that the codeword length $N$ is fixed once training is complete, and designers should select an appropriate length based on their specific performance and application requirements. This limitation could be addressed in future work through more advanced neural network architectures capable of adapting to variable secret lengths.

The neural network architecture, particularly the input and output layer configurations, exhibits strong dependence on the preamble characteristics of the target wireless protocol. This protocol-specific dependency arises from fundamental differences in preamble structure and length across standards. For example, LoRa and BLE protocols employ waveforms with substantially different temporal durations, requiring corresponding architectural adaptations to accommodate these protocol-specific requirements.

\subsection{Secret Message Extraction at Bob Side}\label{sec:Bob}

The radio signal emitted by Alice propagates through the air and is then captured by Bob. It is down-converted to the baseband and subsequently discretized by an ADC. The preamble field of the wireless packet is extracted, denoted as $\mathbf{u}'$. Subsequently, a secret extractor, a de-interleaver, and an FEC decoder are employed to process the preamble waveform, as detailed below.

\subsubsection{Secret Extractor}\label{sec:bits_extractor}

Bob first uses a pre-trained secret extractor to recover the message hidden in the distorted preamble $\mathbf{u}'$, which is mathematically given as
\begin{equation}\label{equ:extractor}
    \mathbf{\hat{c}} = \mathcal{G}_{\mathrm{ext}}(\mathbf{u}';\theta_{\mathrm{ext}}),
\end{equation}
where $\mathcal{G}_{\mathrm{ext}}(\cdot;\theta_{\mathrm{ext}})$ denotes the neural network-based secret extraction process and $\theta_{\mathrm{ext}}$ is the parameter set. It accepts a received preamble $\mathbf{u}'$ and outputs an estimated codeword $\mathbf{\hat{c}}$.  

The architecture of the neural network extractor is shown in Fig.~\ref{fig:nn_architecture}(b), consisting of nine 1D convolutional layers, an adaptive average pooling layer, and a linear layer. First, the received preamble $\mathbf{u}'$ is normalized by dividing by its root mean square (RMS) power, thereby preventing the signal amplitude from affecting the message extraction results. Then the normalized signal is split into real and imaginary parts and fed sequentially into nine 1D convolutional layers to output $N$-channel feature maps. Subsequently, an adaptive average pooling layer and a linear layer are employed to transform the feature map into an $N$-element vector, denoted by $\mathbf{v}$. Finally, the vector $\mathbf{v}$ is processed by a sign function, thereby generating a binary codeword, $\mathbf{\hat{c}}$, where a positive sign denotes 1 and a negative sign denotes 0. The outputs of all convolutional layers are activated by GELU functions.

\subsubsection{De-Interleaver and Forward Error Correction Decoder}

After extracting the codeword $\mathbf{\hat{c}}$ from the packet preamble field, Bob employs the random de-interleaver to reconstruct its original order and then uses an FEC decoder to recover the secret message, which is expressed as
\begin{equation}
    \mathbf{\hat{x}} = \mathcal{F}_{\mathrm{dec}}(\mathbf{\hat{c}}),
\end{equation}
where $\mathcal{F}_{\mathrm{dec}}$ represents the FEC decoding process and $\mathbf{\hat{x}}$ denotes the recovered $M$-bit secret message. The (15, 5) BCH decoder is employed in the LoRa and BLE case studies.

\section{Neural Network Joint Training and Sim-to-Real Fine-Tuning}\label{sec:joint_training}

\begin{figure}[!t]
    \centering
    \includegraphics[width = 3.4in]{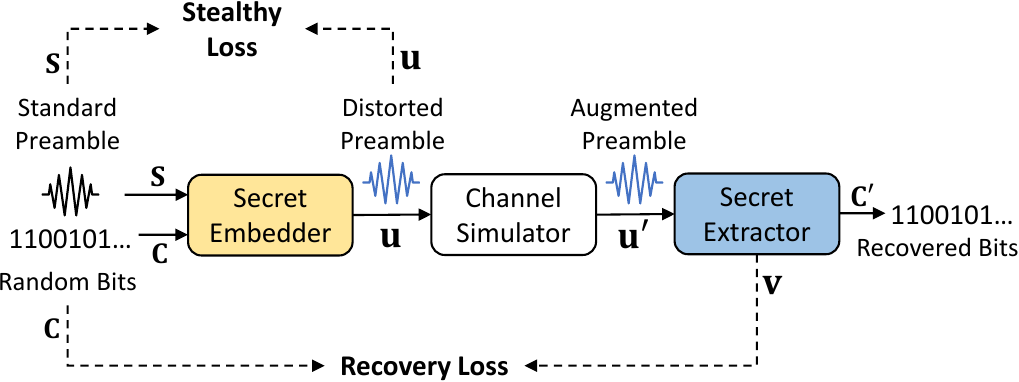}
    \caption{Simulation-driven embedder-extractor joint training scheme.}
    \label{fig:joint_training}
\end{figure}

The neural network–based secret embedder and extractor are fundamental components of the PHY information hiding system. They are jointly trained in a controlled environment, i.e., a physically protected setting, to prevent backdoor, data poisoning, and other training-time attacks~\cite{zhao2024explanation}. After training, the model weights are securely shared between Alice and Bob. An additional sim-to-real fine-tuning stage may be applied to further enhance performance on practical hardware platforms.

\subsection{Training Scheme Overview}

The joint training process is illustrated in Fig.~\ref{fig:joint_training}. It can be observed that the communication process between Alice and Bob is emulated with a secret embedder, a wireless channel simulator, and a secret extractor. At the beginning of each training step, a batch of $N$-bit binary vectors is randomly generated, with the batch size defined as $K$. More specifically, each element in the binary vector is randomly assigned as 0 or 1, i.e., a uniform distribution. After that, a secret embedder is employed to conceal the message $\mathbf{c}$ within the standard preamble $\mathbf{s}$, generating a distorted signal $\mathbf{u}$. Subsequently, a wireless channel simulator is utilized for data augmentation. In particular, it emulates fading, adds noise into $\mathbf{u}$, and outputs an augmented signal $\mathbf{u}'$.
Finally, a secret extractor is employed to recover the embedded binary vector. The following subsections will elaborate on the design of loss functions and the wireless channel simulator.

\subsection{Loss Functions}

Two loss functions are defined to enforce constraints on the neural network training process, namely a recovery loss $\mathcal{L}_{\mathrm{recovery}}$ and a stealthy loss $\mathcal{L}_{\mathrm{stealthy}}$. The recovery loss $\mathcal{L}_{\mathrm{recovery}}$ measures the difference between the embedded random bits and the output vector, which is mathematically given as
\begin{equation}\label{equ:recovery}
    \mathcal{L}_{\mathrm{recovery}} = \frac{1}{K} \sum^{K}_{k=1}\frac{\left \| \mathbf{c}_k - \mathbf{v}_k \right \|_2}{N},
\end{equation}
where $\mathbf{c}_k$ and $\mathbf{v}_k$ represent the embedded random bits and output vector for the $k$-th sample in the training batch, respectively. 
Specifically, the vector $\mathbf{v}$ denotes the neural network output before the sign function, as illustrated in Fig.~\ref{fig:nn_architecture}. 
The symbol $\left \| \cdot \right \|_2$ returns the L2 norm. Meanwhile, the stealthy loss $\mathcal{L}_{\mathrm{stealthy}}$ measures the similarity between the standard and distorted preambles, given as 
\begin{equation}
    \mathcal{L}_{\mathrm{stealthy}} = \frac{1}{K} \sum^{K}_{k=1}\frac{\left \| \mathbf{u}_k - \mathbf{s} \right \|_2}{L},
\end{equation}
where $\mathbf{u}_k$ denotes the distorted preamble for the $k$-th training sample. $\mathbf{s}$ refers to the standard preamble introduced in Section~\ref{sec:info_hiding_basics}. The total loss $\mathcal{L}_{\mathrm{total}}$ is then defined as
\begin{equation}
    \mathcal{L}_{\mathrm{total}} = \alpha \cdot \mathcal{L}_{\mathrm{recovery}} + (1-\alpha) \cdot \mathcal{L}_{\mathrm{stealthy}},
\end{equation}
where $\alpha$ is a hyperparameter that adjusts the weights of the two losses. In all experiments, $\alpha$ is empirically set to 0.6. Note that for a loss function comprising only two terms, employing a single hyperparameter to weight them is mathematically equivalent to using two separate weights, as the relative ratio is what matters for optimization. The objective of the neural network training process is to find a set of parameters that minimize the loss function $\mathcal{L}_{\mathrm{total}}$, which is mathematically expressed as
\begin{equation}
    \Theta = \mathop{\argmin}_{\Theta} \sum_{\mathbf{c}\in \mathcal{D}} \mathcal{L}_{\mathrm{total}}(\mathbf{c}; \Theta),
\end{equation}
where $\Theta = \theta_{\mathrm{emb}} \cup \theta_{\mathrm{ext}}$ denotes the parameters of both the secret embedder and extractor. $\mathcal{D}$ represents the sample space of the $N$-bit binary vector $\mathbf{c}$.

In summary, the recovery loss $\mathcal{L}_{\mathrm{recovery}}$ is intended to ensure that the secret message can be successfully embedded and extracted. The stealthy loss $\mathcal{L}_{\mathrm{stealthy}}$ constrains the degree of the waveform distortion, making the covert link undetectable to eavesdroppers.

\subsection{Wireless Channel Simulator}\label{sec:channel_simulator}

The training process incorporates a wireless channel simulator for data augmentation, which can significantly enhance the neural network's resilience against channel effects. We consider an additive white Gaussian noise (AWGN) and slow flat-fading in this work, expressed as
\begin{equation}
    \mathbf{u}' = \mathbf{h} \cdot \mathbf{u} + \mathbf{z},
\end{equation}
where $\mathbf{h}$ is the complex channel coefficient and $\mathbf{z}$ is the additive noise. The coefficient $\mathbf{h}$ is generated using the Rayleigh fading model. Only the phase of $\mathbf{h}$ is simulated, as the amplitude attenuation does not affect the system performance due to the power normalization module in the secret extractor.  The noise vector $\mathbf{z}$ is calculated using a random SNR uniformly distributed in the range [0, 60]~dB.

\subsection{Sim-to-Real Fine-Tuning}\label{sec:fine_tune}

The scheme introduced in previous subsections leverages simulation data to jointly train a pair of secret embedder and extractor. However, due to the inevitable discrepancy between the simulated and experimental data, the model performance often degrades in practical environments. To address this issue, we propose a sim-to-real fine-tuning scheme as a mitigation approach.

\begin{figure}[!t]
    \centering
    \includegraphics[width = 3.4in]{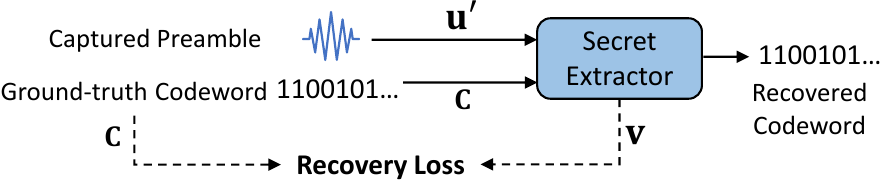}
    \caption{Sim-to-real fine-tuning. The parameters of the secret extractor are fine-tuned with the signals collected by actual transceiver hardware platforms.}
    \label{fig:fine_tuning}
\end{figure}

In essence, the sim-to-real fine-tuning scheme is the use of experimentally collected data to calibrate the parameters $\theta_{\mathrm{ext}}$ of the secret extractor $\mathcal{G}_{\mathrm{ext}}$, which is illustrated in Fig.~\ref{fig:fine_tuning}. This can effectively mitigate performance degradation caused by simulation-to-reality gaps, e.g., hardware imperfections, when the models are deployed on actual transceiver hardware platforms. Firstly, we deploy the secret embedder $\theta_{\mathrm{emb}}$ and extractor $\theta_{\mathrm{ext}}$ on the transmitter and receiver, respectively, and use this hardware system to collect a fine-tuning dataset $\mathcal{D}^\mathrm{tune}$, given as
\begin{equation}
    \mathcal{D}^\mathrm{tune} = \{(\mathbf{u'}_i, \mathbf{c}_i)\}_{i=1}^{I},
\end{equation}
where $I$ denotes the number of samples in the dataset $\mathcal{D}^\mathrm{tune}$. $\mathbf{u'}_i$ is the $i$-th signal collected by the receiver hardware, and $\mathbf{c}_i$ is the corresponding ground-truth codeword. After the collection of $\mathcal{D}^\mathrm{tune}$ is complete, we first input the received signal $\mathbf{u'}$ into the simulation-trained secret extractor $\mathcal{G}_{\mathrm{ext}}$, and then fine-tune the parameter set $\theta_{\mathrm{ext}}$ using the recovery loss $\mathcal{L}_{\mathrm{recovery}}$ defined in (\ref{equ:recovery}).

In summary, the simulation-trained deep learning models tend to experience performance degradation when deployed on practical hardware platforms.
To address this issue, we propose to fine-tune the secret extractor $\mathcal{G}_{\mathrm{ext}}$ using real-collected wireless signals in a controlled short-range, high-SNR LOS setting.
This approach aims to adapt the model to the specific hardware distortions, such as I/Q imbalance and nonlinearity of the power amplifier, rather than to adapt to a particular channel state.
Robustness to time-varying channels is mainly provided by the wireless channel simulator and data augmentation in the joint training stage.
After fine-tuning, the model weights remain fixed and are directly tested in other locations and channel conditions without per-location recalibration.

Note that the sim-to-real fine-tuning is not a mandatory step. The experimental results in Section~\ref{sec:real_world_evaluation} will demonstrate that the simulation-trained neural networks already achieve excellent information hiding performance, and sim-to-real fine-tuning can further enhance the performance. It is crucial to emphasize that the fine-tuning process must be conducted in a controlled environment, which is essential to prevent an attacker from training a surrogate neural network to achieve model stealing~\cite{orekondy2019knockoff}.

\section{Evaluation Metrics}

\textbf{Raw Secret Bit Error Rate:} The raw secret bit error rate (SBER), denoted by $\mathrm{SBER}_{raw}$, measures the difference between the embedded and extracted codewords, i.e., $\mathbf{c}$ and $\mathbf{\hat{c}}$, which is mathematically given as
\begin{equation}
    \mathrm{SBER}_{raw} = \frac{\mathcal{H}(\mathbf{c},\mathbf{\hat{c}})}{N},
\end{equation}
where $\mathcal{H}(\cdot, \cdot)$ calculates the Hamming distance between two binary vectors. The raw SBER metric directly measures the performance of the secret embedder and extractor without the benefit of any error correction technique.


\textbf{Secret Bit Error Rate after FEC:} The FEC-SBER, denoted by $\mathrm{SBER}_{fec}$, measures the difference between the embedded and extracted secret messages, i.e.,  $\mathbf{x}$ and $\mathbf{\hat{x}}$, which is expressed as
\begin{equation}
    \mathrm{SBER}_{fec} = \frac{\mathcal{H}(\mathbf{x},\mathbf{\hat{x}})}{M}.
\end{equation}
$\mathrm{SBER}_{fec}$ is typically lower than $\mathrm{SBER}_{raw}$ since the FEC technique can correct some or all of the errors. 


\textbf{Secret Message Error Rate:} The secret message error rate (SMER) is defined as the ratio between the number of erroneous secret messages $E_{\mathrm{err}}$ and the total number of transmitted secret messages $E_{\mathrm{total}}$, given as
\begin{equation}
    \mathrm{SMER} = \frac{E_{\mathrm{err}}}{E_{\mathrm{total}}}.
\end{equation}
It evaluates the overall quality of the established covert link at the packet level. A high SMER indicates severe secret extraction errors.


\textbf{Packet Loss Rate:} The packet loss rate (PLR) quantifies the reliability of the primary communication link at the packet level. It is defined as the ratio of packets that are either misdetected or fail the cyclic redundancy check (CRC) to the total number of transmitted packets:
\begin{equation}
\mathrm{PLR} = \frac{P_{\mathrm{lost}}}{P_{\mathrm{total}}},
\end{equation}
where $P_{\mathrm{lost}}$ denotes the number of lost packets and $P_{\mathrm{total}}$ is the total number of transmitted packets. A packet is regarded as lost if the receiver fails to synchronize with the preamble or if the header or payload decoding fails the CRC validation.
Importantly, the PLR reflects the impact of the preamble-based secret embedding on the primary communication link and does not evaluate the performance of the covert channel.

\section{Simulation Evaluation}\label{sec:simulation_evaluation}

\subsection{Simulation Configuration}\label{sec:simulation_config}

\subsubsection{LoRa and BLE PHY Configuration}

The LoRa PHY is implemented in MATLAB, building upon the existing reverse-engineering study~\cite{xu2023demodulation}. The spreading factor and bandwidth are set to 7 and 500~kHz, respectively. The sampling rate is configured to 1~MHz. In this PHY configuration, the LoRa preamble field contains 2,048 I/Q samples, which is visualized in Fig.~\ref{fig:preamble_waveform}(a).

The BLE PHY is developed based on the official MATLAB Bluetooth toolbox~\footnote{https://www.mathworks.com/products/bluetooth.html}. The BLE PHY mode is set to LE1M, where the packet preamble is a `10101010' sequence. Each BLE symbol contains 16 I/Q samples. In this configuration, the preamble waveform contains 128 I/Q samples, which is visualized in Fig.~\ref{fig:preamble_waveform}(d).

In both LoRa and BLE cases, we simulate 5,000 independent wireless packets and record the $\mathrm{SBER}_{raw}$, $\mathrm{SMER}$, and $\mathrm{PLR}$ for subsequent analysis.

\subsubsection{Neural Network Configuration}\label{sec:nn_config_simulation}

The neural networks are implemented with the PyTorch library. The secret embedder and extractor are trained in pairs as described in Section~\ref{sec:joint_training}. The RMSprop optimizer is employed during the training process with a batch size of 64. The initial learning rate is set to 0.0003, and if the validation loss does not decrease within 10 epochs, the current learning rate is reduced by a factor of 0.1, i.e., a reduce-on-plateau scheduler. Training stops when the validation loss does not decrease for 20 epochs.

\begin{table*}[!t]
  \centering
  \caption{Simulation results of PHY information hiding for LoRa and BLE protocols.}
  \begin{tabular}{cclcccccccccc}
    \toprule
    \multirow{2}[4]{*}{Protocol} & \multirow{2}[4]{*}{Secret Length} & \multicolumn{1}{c}{\multirow{2}[4]{*}{Metrics}} & \multicolumn{9}{c}{SNR} \\
    \cmidrule{4-12}
    & & & -10~dB & -5~dB & 0~dB & 5~dB & 10~dB & 15~dB & 20~dB & 25~dB & 30~dB \\
    \midrule
    \multirow{10}{*}{LoRa} 
      & \multirow{3}[2]{*}{60 bit} 
      & $\mathrm{SBER}_{raw}$ & 0.4851 & 0.3776 & 0.1445 & 0.0174 & 0.0022 & 0.0017 & 0.0017 & 0.002 & 0.0019 \\
      & & $\mathrm{SMER}$ & 1 & 0.9992 & 0.447 & 0.0028 & 0 & 0 & 0 & 0 & 0 \\
      & & $\mathrm{PLR}$ & 0.6406 & 0.0038 & 0 & 0 & 0 & 0 & 0 & 0 & 0 \\
    \cmidrule{2-12}
      & \multirow{3}[2]{*}{120 bit} 
      & $\mathrm{SBER}_{raw}$ & 0.4865 & 0.4193 & 0.2237 & 0.0631 & 0.0246 & 0.0175 & 0.0163 & 0.0162 & 0.0159 \\
      & & $\mathrm{SMER}$ & 1 & 1 & 0.97 & 0.1036 & 0.007 & 0.004 & 0.0028 & 0.0022 & 0.0022 \\
      & & $\mathrm{PLR}$ & 0.6882 & 0.0046 & 0 & 0 & 0 & 0 & 0 & 0 & 0 \\
    \cmidrule{2-12}
      & \multirow{3}[2]{*}{180 bit} 
      & $\mathrm{SBER}_{raw}$ & 0.4933 & 0.4424 & 0.2678 & 0.1144 & 0.0626 & 0.0513 & 0.0478 & 0.0467 & 0.0466 \\
      & & $\mathrm{SMER}$ & 1 & 1 & 0.9996 & 0.6156 & 0.1394 & 0.0814 & 0.0724 & 0.0628 & 0.0618 \\
      & & $\mathrm{PLR}$ & 0.6308 & 0.0088 & 0 & 0 & 0 & 0 & 0 & 0 & 0 \\
    \cmidrule{2-12}
      & No secret & $\mathrm{PLR}$ & 0.4978 & 0.0034 & 0 & 0 & 0 & 0 & 0 & 0 & 0 \\
    \midrule
    \multirow{10}{*}{BLE} 
      & \multirow{3}[2]{*}{60 bit} 
      & $\mathrm{SBER}_{raw}$ & 0.4729 & 0.4011 & 0.2621 & 0.1116 & 0.0378 & 0.0173 & 0.0135 & 0.0116 & 0.0112 \\
      & & $\mathrm{SMER}$ & 1 & 1 & 0.9447 & 0.2606 & 0.012 & 0.0024 & 0.0018 & 0.0016 & 0.0006 \\
      & & $\mathrm{PLR}$ & 0.9762 & 0.0764 & 0.0004 & 0 & 0 & 0 & 0 & 0 & 0 \\
    \cmidrule{2-12}
      & \multirow{3}[2]{*}{120 bit} 
      & $\mathrm{SBER}_{raw}$ & 0.4785 & 0.4333 & 0.3378 & 0.2139 & 0.1221 & 0.079 & 0.0651 & 0.0599 & 0.0591 \\
      & & $\mathrm{SMER}$ & 1 & 1 & 1 & 0.9726 & 0.5428 & 0.1986 & 0.123 & 0.1026 & 0.0974 \\
      & & $\mathrm{PLR}$ & 0.9796 & 0.068 & 0.0006 & 0 & 0 & 0 & 0 & 0 & 0 \\
    \cmidrule{2-12}
      & \multirow{3}[2]{*}{180 bit} 
      & $\mathrm{SBER}_{raw}$ & 0.4927 & 0.4874 & 0.4581 & 0.4173 & 0.3926 & 0.3839 & 0.3806 & 0.3795 & 0.3791 \\
      & & $\mathrm{SMER}$ & 1 & 1 & 1 & 1 & 1 & 1 & 1 & 1 & 1 \\
      & & $\mathrm{PLR}$ & 0.9876 & 0.0666 & 0.0008 & 0 & 0 & 0 & 0 & 0 & 0 \\
    \cmidrule{2-12}
      & No secret & $\mathrm{PLR}$ & 0.9752 & 0.064 & 0.0002 & 0 & 0 & 0 & 0 & 0 & 0 \\
    \bottomrule
  \end{tabular}%
  \label{tab:simulation_results_combined}%
\end{table*}

\subsection{Simulation Results of LoRa Information Hiding}

\subsubsection{Secret Extraction Performance}

The simulation results for LoRa information hiding are summarized in Table~\ref{tab:simulation_results_combined}, which presents system performance across varying secret lengths and SNR conditions. The secret embedding and extraction performance metrics, $\mathrm{SBER}_{raw}$ and $\mathrm{SMER}$, demonstrate a clear dependence on both SNR and secret length.

At SNR levels below 5 dB, both $\mathrm{SBER}_{raw}$ and $\mathrm{SMER}$ remain prohibitively high, rendering reliable secret extraction infeasible. However, performance improves substantially once SNR exceeds 10 dB. For 60-bit and 120-bit secrets, $\mathrm{SBER}_{raw}$ drops below 0.025 and $\mathrm{SMER}$ decreases to less than 0.01, indicating that nearly all embedded secret messages can be accurately recovered. While the 180-bit configuration remains functional, it exhibits higher error rates (e.g., $SMER \approx 0.14$ at 10 dB). This degradation is expected, as longer secrets place greater demands on the neural network's capacity. Adopting more advanced network architectures would likely mitigate these limitations and enhance robustness for longer secret lengths.

In summary, the simulation results demonstrate that the proposed LoRa PHY information hiding approach is both effective and practical when SNR exceeds 10 dB, with secret lengths up to 120 bits being reliably embedded and extracted. These findings validate the feasibility of the proposed PHY-layer information hiding technique and highlight the inherent trade-off between secret throughput and extraction reliability.

\subsubsection{Impact on the Primary LoRa Link}

The PLR results in Table~\ref{tab:simulation_results_combined} confirm that secret embedding does not degrade the primary communication link. When SNR exceeds 0~dB, the PLR is identical to the no-secret baseline, indicating that the introduced preamble perturbations do not impair packet detection, synchronization, or payload decoding at all. 
This demonstrates that packet losses are primarily governed by channel noise rather than the proposed PHY-layer information hiding mechanism.

Furthermore, even at low SNR levels below 0~dB, the difference in PLR between the embedded and baseline scenarios remains negligible, demonstrating robust coexistence between the covert channel and the primary LoRa transmission. This preserved link reliability shows the feasibility of leveraging preamble manipulation for secure information hiding without compromising the integrity of the main communication link. These findings are further validated through experiments in Section~\ref{sec:compatiblity_lora}, where a commercial LoRa chipset successfully received secret-embedded LoRa packets.

\subsubsection{Impact on LoRa CFO Estimation Errors}\label{sec:impact_cfo_lora}

\begin{figure}[!t]
	\centering
	\subfloat[]{\includegraphics[width=1.7in]{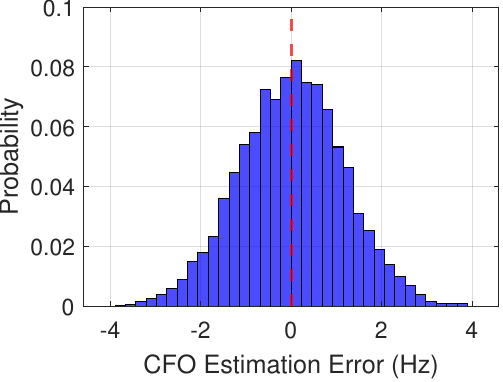}}
	\subfloat[]{\includegraphics[width=1.7in]{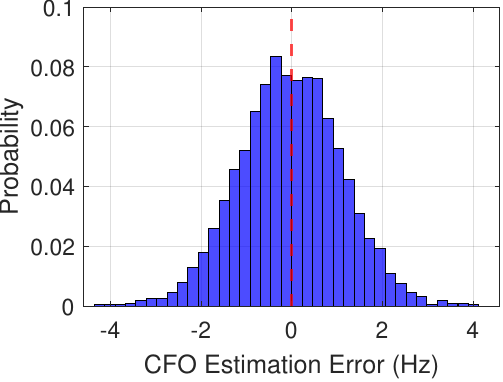}}
	\caption{Distribution of CFO estimation errors across 5,000 simulated LoRa transmissions. Autocorrelation-based CFO estimation algorithm is utilized, and the red dashed line marks zero CFO estimation error. (a) With the standard preamble. (b) With a secret-embedded preamble.}
	\label{fig:cfo_errors_lora}
\end{figure}
In addition to packet detection, the packet preamble serves as a critical component for carrier frequency offset (CFO) estimation. The LoRa PHY employed in this work adopts an autocorrelation-based CFO estimation algorithm. As discussed in Section~\ref{sec:motivation_challenges}, since the repetition structure of the preamble is preserved, autocorrelation-based algorithms are inherently unaffected.

To validate this claim, we conducted MATLAB simulations involving 5,000 LoRa transmissions, in which random CFOs were injected into the transmitted LoRa waveforms. The CFO values are uniformly distributed in the range of [-10~ppm, 10~ppm], i.e., $\pm 4.33$~kHz for a center frequency of 433~MHz. At the receiver, the CFO was estimated, and the estimation error was computed as the difference between the injected and estimated CFO values. The simulations were performed twice, with PHY information hiding enabled and disabled, respectively.

The resulting CFO estimation errors are illustrated as histograms in Fig.~\ref{fig:cfo_errors_lora}. The results show negligible performance differences between the standard preamble and the secret-embedded preamble for CFO estimation. In both cases, the CFO estimation error is tightly bounded within 
[-4~Hz, 4~Hz], demonstrating high estimation accuracy. These results confirm that the proposed PHY information hiding scheme does not degrade the performance of the autocorrelation-based CFO estimation algorithm.

\subsection{Simulation Results of BLE Information Hiding}

\subsubsection{Secret Extraction Performance}
The simulation results for BLE information hiding are presented in Table~\ref{tab:simulation_results_combined}. Similar to the LoRa results, $\mathrm{SBER}_{raw}$ and $\mathrm{SMER}$ in BLE exhibit strong dependence on both SNR and secret length, with reliable secret extraction achievable only when SNR exceeds 10 dB. Across all tested secret lengths (60, 120, and 180 bits), $\mathrm{SMER}$ remains close to 1 below 5 dB, rendering covert communication practically infeasible in low-SNR regimes.
Performance improves dramatically with increasing SNR; however, the maximum reliable secret length is considerably lower than in LoRa. For the most robust 60-bit configuration, once SNR reaches 10 dB, $\mathrm{SBER}_{raw}$ drops below 0.038 and $\mathrm{SMER}$ falls to 0.012. Above 15 dB, $\mathrm{SMER}$ approaches zero, enabling near-perfect message recovery. In contrast, longer secrets (120 bits and 180 bits) exhibit significantly higher error floors and require high SNR for acceptable $\mathrm{SMER}$, indicating that the current BLE embedding scheme reliably supports up to 60 bits under typical SNR conditions. 

The performance gap between LoRa and BLE is expected due to their differing temporal resolutions: at our sampling rate, LoRa preambles contain 2,048 I/Q samples, while BLE’s are limited to just 128 I/Q samples. This orders-of-magnitude difference severely restricts the available space for secret embedding in BLE preambles, directly limiting both capacity and robustness. Despite these constraints, the 60-bit performance demonstrates that reliable secret embedding remains feasible when SNR exceeds 10~dB.

\subsubsection{Impact on the Primary BLE Link}

The results in Table~\ref{tab:simulation_results_combined} demonstrate that the primary BLE communication link remains unaffected by the secret embedding process. The PLR results for all secret lengths closely match the no-secret baseline across all SNR levels. When SNR exceeds 0 dB, PLR in both baseline and secret-embedded scenarios reduces to zero, demonstrating that preamble perturbations do not interfere with BLE synchronization or payload decoding at all.

Even in low-SNR regimes, the PLR difference between the two cases is minimal, confirming that packet losses are dominated by channel noise rather than the PHY information hiding mechanism. These results demonstrate that the proposed PHY-layer embedding technique can coexist with the BLE primary link without compromising link reliability. This is further validated experimentally in Section~\ref{sec:compatibility_smartphone}, where an Android smartphone successfully received secret-embedded BLE packets.

\subsubsection{Impact on BLE CFO Estimation Errors}\label{sec:impact_cfo_ble}

\begin{figure}[!t]
	\centering
	\subfloat[]{\includegraphics[width=1.7in]{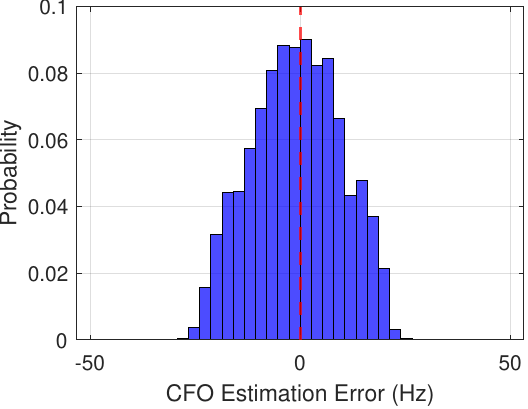}}
	\subfloat[]{\includegraphics[width=1.7in]{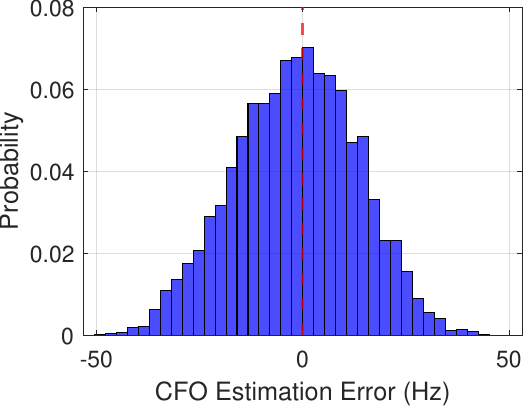}}
	\caption{Distribution of CFO estimation errors across 5,000 simulated BLE transmissions. FFT-based CFO estimation algorithm is utilized, and the red dashed line marks zero CFO estimation error. (a) With the standard preamble. (b) With a secret-embedded preamble.}
	\label{fig:cfo_errors_ble}
\end{figure}

We conducted simulations analogous to those in Section~\ref{sec:impact_cfo_lora} to further verify that the proposed PHY information hiding scheme does not compromise CFO estimation performance. Specifically, 5,000 BLE transmissions were simulated, with the true CFO values uniformly distributed over the range 
[-10~ppm, 10~ppm], corresponding to 
$\pm24$~kHz at a carrier frequency of 2.4~GHz.

The resulting CFO estimation errors are illustrated in Fig.~\ref{fig:cfo_errors_ble}. As observed, the use of the secret-embedded preamble leads to a slight degradation in CFO estimation accuracy. In particular, the estimation error range increases from approximately 
$\pm30$~Hz to $\pm50$~Hz. This result differs from the LoRa case discussed in Section~\ref{sec:impact_cfo_lora}, where PHY information hiding has no observable impact on CFO estimation. The discrepancy arises because the BLE PHY implementation relies on an FFT-based CFO estimation algorithm, rather than an autocorrelation-based approach.

Nevertheless, the secret-embedded preamble remains highly similar to the standard preamble, resulting in only an increase of about 20~Hz in the CFO estimation error. This corresponds to approximately 0.0083 ppm at 2.4~GHz and remains well within the tolerance of practical wireless communication systems.

\section{Experiment Evaluation}\label{sec:real_world_evaluation}

\subsection{Experiment Configurations}

\subsubsection{Hardware Prototype}

To successfully deploy a PHY information hiding system, the transceivers must support the execution of neural networks and provide access to physical layer I/Q samples. Furthermore, the transmitter should be able to send customized radio waveforms. It is worth emphasizing that the proposed technique does not inherently require an SDR-based implementation. Its essential requirement is the ability to manipulate baseband I/Q samples at the transmitter and access baseband I/Q samples at the receiver. SDR technology is adopted in this work mainly as a feasibility-verification prototype, because it decouples software and hardware and provides flexible waveform generation and I/Q sample access for algorithm validation. SDR has been extensively employed in satellite, defense, drone, and 5G cellular applications, making the proposed PHY information hiding readily verifiable on practical hardware platforms.

Two USRP B210 SDR~\cite{usrpb210url} platforms are employed as wireless transmitter and receiver in the experiments. They are connected to two battery-powered laptops via USB ports, which run MATLAB and Python programs to perform LoRa/BLE transmission and PHY information hiding tasks. We emphasize that this prototype is incapable of operating in real time due to the limited computing capabilities of the laptop CPUs. In practical deployments, the secret embedder and extractor can be integrated into the digital baseband processing chain and implemented on FPGA, DSP, dedicated neural accelerators, or eventually baseband chips to achieve real-time performance with substantially lower cost and power consumption.

It is important to note that no external RF amplifier is used in the experiment. For both protocols, we employ standard omnidirectional antennas: a 433~MHz antenna with 4~dBi gain for LoRa and a 2.4~GHz antenna with 2~dBi gain for BLE. Therefore, the measured communication distance is jointly determined by the USRP B210 transmit power and the antenna characteristics, rather than reflecting a fundamental limit of the proposed method. For this reason, we adopt SNR as the primary evaluation metric throughout the paper, since it provides a more hardware-independent measure of system performance.

\subsubsection{LoRa and BLE PHY Configuration}

The LoRa and BLE PHY are identical to the configurations used in simulation, which is introduced in Section~\ref{sec:simulation_config}. The SDR transmitter continuously broadcasts LoRa or BLE signals in the experiments. The secret message is randomly generated at the transmitter side and sent as the packet payload. The receiver parses the payload to obtain the ground-truth secret message, which is used to calculate the evaluation metrics.

For LoRa, the secret message length $M$ is fixed at 40 bits unless otherwise specified, resulting in a 120-bit codeword length $N$. For BLE, the secret message length $M$ is fixed at 20 bits unless otherwise specified, corresponding to a 60-bit codeword embedded in the preamble field. Unless otherwise specified, 500 BLE and LoRa packets are collected at each location. The evaluation metrics and estimated SNR values are provided as the average of all collected packets. The SNR of each packet is estimated by computing the ratio between the average power of the received packet waveform and that of the noise-only I/Q samples preceding the packet.

Unless otherwise specified, each BER and SMER value is averaged over successfully received packets at the corresponding location. In total, we collect 4,000 LoRa packets across 8 locations and 2,000 BLE packets across 4 locations, with each packet carrying a 120-bit and 60-bit codeword, respectively. Although the USB-based USRP B210 prototype may occasionally introduce dropped samples or packet losses, the averaged results over thousands of packets provide representative empirical evidence for the feasibility of the proposed PHY-layer information hiding method under the evaluated prototype conditions.

\subsubsection{Neural Network Configuration}

It is worth noting that the neural networks deployed in real-world experiments are trained exclusively on simulation data unless otherwise specified. In other words, the models are identical to those used in Section~\ref{sec:simulation_evaluation}, with the same architectures and weights. Once deployed in practical environments, the neural networks can be fine-tuned using real-world collected data to further enhance performance. The effects of fine-tuning for LoRa and BLE are demonstrated in Sections~\ref{sec:effect_fine_tune_lora} and~\ref{sec:effect_fine_tune_ble}, respectively.



\subsection{Experiment Results of LoRa Information Hiding}

\subsubsection{Secret Extraction Performance in Indoor Environment}\label{sec:indoor_evaluation}

\begin{figure}[!t]
    \centering
    \includegraphics[width = 3.4in]{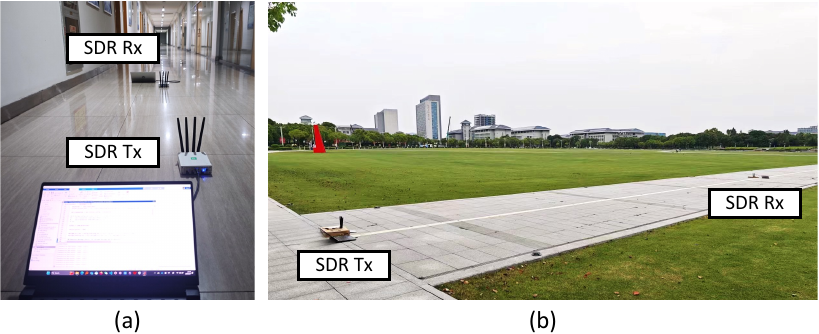}
    \caption{Experiment environments. (a) Indoor environment. (b) Outdoor environment.}
    \label{fig:experiment_pic}
\end{figure}

\begin{figure}[!t]
    \centering
    \includegraphics[width = 3in]{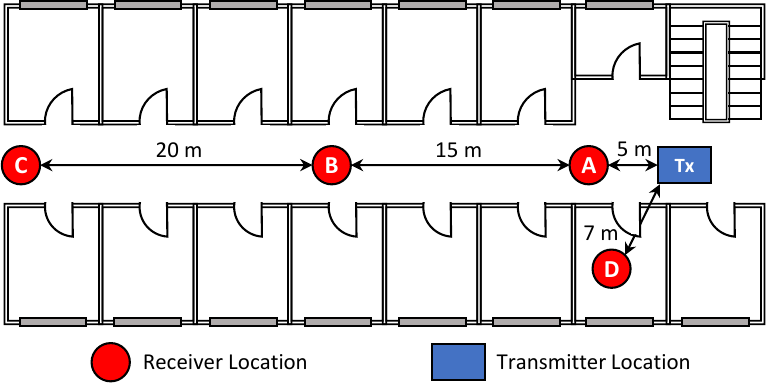}
    \caption{Floor plan of the office building. }
    \label{fig:floorplan}
\end{figure}

\begin{table}[!b]
  \centering
  \caption{Experimental results of LoRa information hiding in the indoor environment. 120 bits are embedded within the preamble field.}
    \begin{tabular}{cclrrr}
    \toprule
    \multirow{2}[4]{*}{Condition} & \multirow{2}[4]{*}{Loc.} & \multicolumn{1}{c}{\multirow{2}[4]{*}{Avg. SNR}} & \multicolumn{3}{c}{Metrics} \\
\cmidrule{4-6}          &       &       & \multicolumn{1}{l}{$\mathrm{SBER}_{raw}$} & \multicolumn{1}{l}{$\mathrm{SBER}_{fec}$} & \multicolumn{1}{l}{$\mathrm{SMER}$} \\
    \midrule
    \multirow{3}[2]{*}{LOS} & A     & 26.39 dB & 0.0242 & 0.0002 & 0.0039 \\
          & B     & 13.37 dB & 0.0423 & 0.0021 & 0.0456 \\
          & C     & 6.65 dB & 0.1595 & 0.0745 & 0.747 \\
    \midrule
    NLOS  & D     & 15.07 dB  & 0.0464 & 0.0021 & 0.0419 \\
    \bottomrule
    \end{tabular}%
  \label{tab:indoor_results}%
\end{table}%

We then carried out an indoor experimental evaluation in an office building. The photos of the experimental environments and the floor plan are given in Fig.~\ref{fig:experiment_pic}(a) and Fig.~\ref{fig:floorplan}, respectively. As depicted in the figures, the SDR transmitter remained stationary while the receiver was sequentially placed at locations A, B, C, and D. Specifically, locations A, B, and C are positioned along a typical corridor, whereas location D is situated inside a standard office room. The environment contains common office furniture, including chairs, desks, bookshelves, and computer monitors. All experiments were performed with the door closed, and the walls are made of concrete.

The experimental results are presented in Table~\ref{tab:indoor_results}. It can be observed that the $\mathrm{SMER}$ is lower than 0.05 when the receiver is placed at locations A, B, and D, indicating a promising PHY information hiding performance. However, the $\mathrm{SMER}$ at location C is 0.747, which implies that most covert messages cannot be extracted without errors and may need retransmission. The high $\mathrm{SMER}$ can be attributed to the low SNR of 6.65~dB on average. Moreover, it can be seen that the $\mathrm{SMER}$ for location D is 0.0419, which indicates that the designed LoRa information hiding system can achieve excellent performance even in non-line-of-sight (NLOS) scenarios.

\subsubsection{Secret Extraction Performance in Outdoor Environment}\label{sec:outdoor_evaluation}

\begin{table}[!t]
  \centering
  \caption{Experimental results of LoRa information hiding in the outdoor environment. 120 bits are embedded within the preamble field.}
    \begin{tabular}{crrrr}
    \toprule
    \multirow{2}[4]{*}{Distance} & \multicolumn{1}{c}{\multirow{2}[4]{*}{Avg. SNR}} & \multicolumn{3}{c}{Metrics} \\
\cmidrule{3-5}          &       & \multicolumn{1}{l}{$\mathrm{SBER}_{raw}$} & \multicolumn{1}{l}{$\mathrm{SBER}_{fec}$} & \multicolumn{1}{l}{$\mathrm{SMER}$} \\
    \midrule
    10 m  & 21.17 dB & 0.0363 & 0.0012 & 0.0219 \\
    20 m  & 11.55 dB& 0.0558 & 0.0068 & 0.0931 \\
    30 m  & 8.1 dB& 0.1073 & 0.0441 & 0.3713 \\
    40 m  & 5.25 dB& 0.1698 & 0.0951 & 0.7386 \\
    \bottomrule
    \end{tabular}%
  \label{tab:outdoor_results}
\end{table}%

As shown in Fig.~\ref{fig:experiment_pic}(b), outdoor evaluation was carried out in a roadside open field with vehicles passing by, including cars and bicycles. Vehicles passed by approximately every 20--40~seconds. Each passing vehicle could introduce rapid fluctuations to the wireless channel, including abrupt variations in the SNR.

The SDR transmitter is kept stationary, while the receiver is placed at four distances of 10~m, 20~m, 30~m, and 40~m. The experimental results in Table~\ref{tab:outdoor_results} demonstrate that the system performance decreases as the distance increases. Specifically, the $\mathrm{SMER}$ increases from 0.0219 to 0.7386, and the $\mathrm{SBER}_{raw}$ rises from 0.0363 to 0.1698 when the transceiver distance increases from 10~m to 40~m. This performance reduction can be attributed to the deterioration in signal quality, as the estimated SNR gradually declines from 21.17~dB to 5.25~dB.

It is important to note that the relatively short outdoor communication range, i.e., up to 40~m, is a consequence of the constrained transmission power of the USRP B210. Therefore, instead of using the absolute communication distance, we adopt SNR as the evaluation metric, as it offers a more general and transferable measure for assessing communication performance across different hardware settings.

\subsubsection{Effect of Sim-to-Real Fine-Tuning}\label{sec:effect_fine_tune_lora}

\begin{table}[!t]
  \centering
  \caption{Effect of sim-to-real fine-tuning in LoRa information hiding. The secret extractor is fine-tuned using the signals collected at location A in the indoor environment and tested on the others.}
    \begin{tabular}{clrrr}
    \toprule
    \multirow{2}[4]{*}{Dataset} & \multicolumn{1}{c}{\multirow{2}[4]{*}{Model}} & \multicolumn{3}{c}{Metrics} \\
    \cmidrule{3-5}      & \multicolumn{1}{c}{} & \multicolumn{1}{l}{$\mathrm{SBER}_{raw}$} & \multicolumn{1}{l}{$\mathrm{SBER}_{fec}$} & \multicolumn{1}{l}{$\mathrm{SMER}$} \\
    \midrule
    Indoor & \multicolumn{1}{l}{Simulation} & 0.0242 & 0.0002 & 0.0039 \\
    Location A & \multicolumn{1}{l}{Fine-tuned} & \multicolumn{1}{c}{-} & \multicolumn{1}{c}{-} & \multicolumn{1}{c}{-} \\
    \midrule
    Indoor & \multicolumn{1}{l}{Simulation} & 0.0423 & 0.0021 & 0.0456 \\
    Location B & \multicolumn{1}{l}{Fine-tuned} & 0.037 & 0.0009 & 0.0198 \\
    \midrule
    Indoor & \multicolumn{1}{l}{Simulation} & 0.1595 & 0.0745 & 0.747 \\
    Location C & \multicolumn{1}{l}{Fine-tuned} & 0.1337 & 0.0494 & 0.6215 \\
    \midrule
    Indoor & \multicolumn{1}{l}{Simulation} & 0.0464 & 0.0021 & 0.0419 \\
    Location D & \multicolumn{1}{l}{Fine-tuned} & 0.0443 & 0.0017 & 0.0399 \\
    \midrule
    Outdoor & Simulation & 0.0363 & 0.0012 & 0.0219 \\
    10 m  & Fine-tuned & 0.0319 & 0.0003 & 0.0099 \\
    \midrule
    Outdoor & Simulation & 0.0558 & 0.0068 & 0.0931 \\
    20 m  & Fine-tuned & 0.0519 & 0.0044 & 0.0634 \\
    \midrule
    Outdoor & Simulation & 0.1073 & 0.0441 & 0.3713 \\
    30 m  & Fine-tuned & 0.095 & 0.0319 & 0.3034 \\
    \midrule
    Outdoor & Simulation & 0.1698 & 0.0951 & 0.7386 \\
    40 m  & Fine-tuned & 0.1403 & 0.0642 & 0.6376 \\
    \bottomrule
    \end{tabular}%

  \label{tab:fine-tune}%
\end{table}%

The neural networks employed in the above subsections are trained entirely using the simulation data. However, their performance inevitably degrades when deployed in practical environments because of the simulation-to-reality gap, which can be concluded by comparing the results in Table~\ref{tab:simulation_results_combined} and Table~\ref{tab:indoor_results}. The sim-to-real fine-tuning introduced in Section~\ref{sec:fine_tune} serves as an effective mitigation strategy, which is experimentally validated.

Specifically, the simulation-trained secret extractor is fine-tuned using the signals collected at location A in the indoor building, which is a short-range LOS setting with an average SNR of 26.39~dB, and tested on the rest of the real-collected datasets. The initial learning rate and batch size are configured as 0.0001 and 32, respectively. The results are illustrated in Table~\ref{tab:fine-tune}. It can be observed that the fine-tuned model achieves improved performance in comparison to the simulation-trained one. For instance, when tested at location B, fine-tuning reduces the $\mathrm{SBER}_{fec}$ and $\mathrm{SMER}$ from 0.0021 to 0.0009, and 0.0456 to 0.0198, respectively, almost half the original value. The results on outdoor datasets also show significant improvement, with the $\mathrm{SMER}$ being reduced by more than 0.1 when the transceivers are placed 40~m apart. In conclusion, the proposed sim-to-real fine-tuning can effectively mitigate the performance degradation caused by the simulation-to-reality gap.

\subsubsection{Compatibility with Commodity LoRa Chips}\label{sec:compatiblity_lora}

\begin{figure}[!t]
    \centering
    \includegraphics[width = 3.1in]{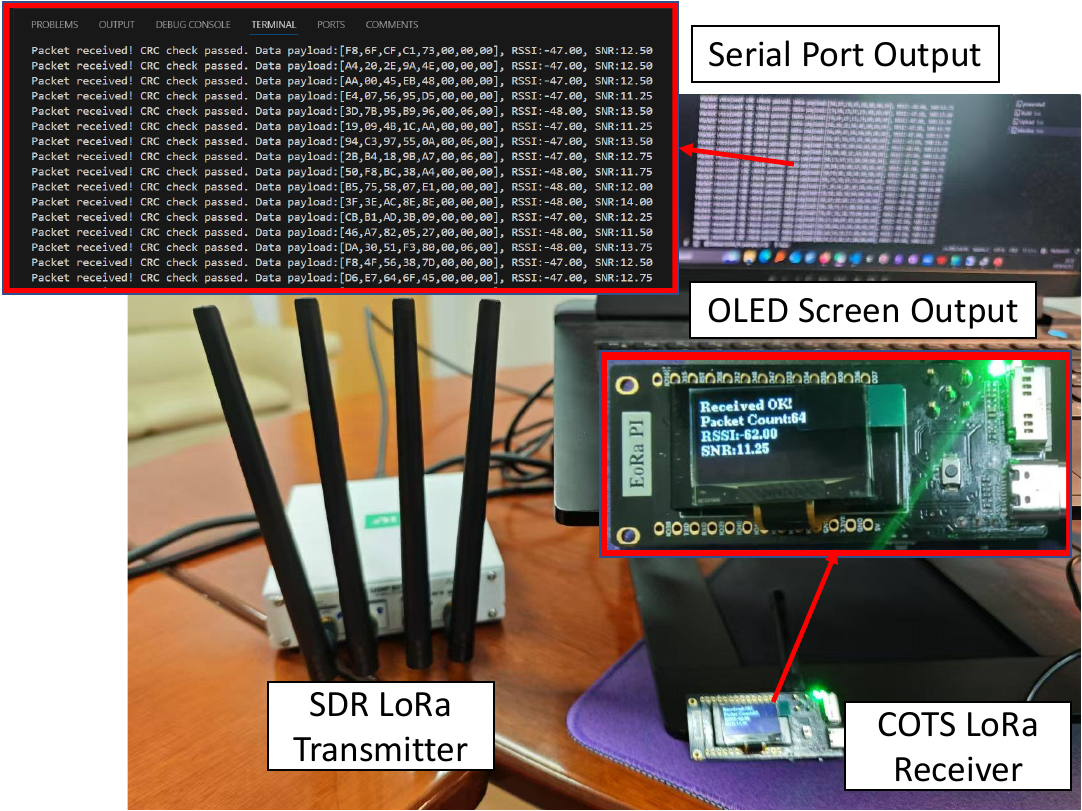}
    \caption{LoRa signal reception using a commodity receiver. The transmitter is an SDR running PHY information hiding and LoRa PHY program. The receiver is a COTS LoRa development board, and the receive status is output via an OLED screen and serial port.}
    \label{fig:cots_reception}
\end{figure}

The discussion in Section~\ref{sec:motivation_challenges} claims that the intentionally introduced physical layer distortion does not affect the normal packet reception and synchronization process. This is experimentally validated through the utilization of a commercial-off-the-shelf LoRa device to receive the signal transmitted from an SDR. As shown in Fig.~\ref{fig:cots_reception}, the transmitter is a USRP B210 SDR running PHY information hiding and LoRa PHY programs, i.e., 60~bits are embedded within the preamble waveform before transmission. Subsequently, a commodity LoRa development board, the EoRa-S3-400TB, which incorporates a Semtech SX1268 chip, is employed for signal reception. It can be observed from the board screen and serial port outputs that the LoRa packets are successfully received. The cyclic redundancy checks are passed, indicating that the payload contents have been decoded without errors.

The experimental results demonstrate that the proposed PHY information hiding technique does not affect the primary communication link. Furthermore, as most commercial RF chips do not provide access to waveform-level information, i.e., I/Q samples, the eavesdroppers have to be equipped with a dedicated device in order to detect the existence of the covert channel, which increases the cost of eavesdropping.

\subsection{Experiment Results of BLE Information Hiding}

\subsubsection{Performance of BLE Information Hiding}

\begin{table}[!t]
  \centering
  \caption{Experimental results of BLE information hiding. 60 bits are embedded within the preamble field.}
    \begin{tabular}{clrrr}
    \toprule
    \multirow{2}[4]{*}{Distance} & \multicolumn{1}{c}{\multirow{2}[4]{*}{Avg. SNR}} & \multicolumn{3}{c}{Metrics} \\
\cmidrule{3-5}          &       & \multicolumn{1}{l}{$\mathrm{SBER}_{raw}$} & \multicolumn{1}{l}{$\mathrm{SBER}_{fec}$} & \multicolumn{1}{l}{$\mathrm{SMER}$} \\
    \midrule
    2 m   & 22.54 dB & 0.0405 & 0.0018 & 0.0159 \\
    5 m   & 12.87 dB & 0.0627 & 0.0072 & 0.0659 \\
    7~m    & 10.88 dB & 0.0784 & 0.0131 & 0.1198 \\
    10~m   & 5.73 dB & 0.1907 & 0.1337 & 0.7086 \\
    \bottomrule
    \end{tabular}%
    \label{tab:exp_results_ble}%
\end{table}%

We then carried out experiments to validate the system design. The experiments are conducted within the same office building illustrated in Section~\ref{sec:indoor_evaluation}, with the floor plan presented in Fig.~\ref{fig:floorplan}. The transmitter is fixed, while the receiver is moved along the corridor to 2~m, 5~m, 7~m, and 10~m apart. The experimental results are given in Table~\ref{tab:exp_results_ble}, which demonstrate that a high-quality BLE covert link can be established when the SNR is above 10~dB. Specifically, the $\mathrm{SMER}$ is as low as 0.0159, 0.0659, and 0.1198, respectively, when the transceivers are placed 2~m, 5~m, and 7~m apart. However, the $\mathrm{SMER}$ increases significantly to 0.7086 when the SNR is around 5.73~dB.
Note that typical indoor BLE links routinely operate in the 10--20~dB SNR range, which aligns well with the reliable extraction threshold observed above. Although the SNR required for reliable secret extraction may be slightly higher than that for normal BLE packet reception, the proposed PHY information hiding scheme is still feasible in typical scenarios.
As noted earlier in Section~\ref{sec:outdoor_evaluation}, the limited communication distance is due to the low transmission power of the USRP B210 board, and an external power amplifier can bring about significant improvement.

\subsubsection{Effect of Sim-to-Real Fine-Tuning}\label{sec:effect_fine_tune_ble}

\begin{table}[!t]
  \centering
  \caption{Effect of sim-to-real fine-tuning in BLE information hiding. The secret extractor is fine-tuned at a 2~m distance and tested on the others.}
    \begin{tabular}{ccrrr}
    \toprule
    \multirow{2}[4]{*}{Dataset} & \multirow{2}[4]{*}{Model} & \multicolumn{3}{c}{Metrics} \\
    \cmidrule{3-5}      &       & \multicolumn{1}{l}{$\mathrm{SBER}_{raw}$} & \multicolumn{1}{l}{$\mathrm{SBER}_{fec}$} & \multicolumn{1}{l}{$\mathrm{SMER}$} \\
    \midrule
    \multirow{2}[2]{*}{2 m} & Simulation & 0.0405 & 0.0018 & 0.0159 \\
          & Fine-tuned & \multicolumn{1}{c}{-} & \multicolumn{1}{c}{-} & \multicolumn{1}{c}{-} \\
    \midrule
    \multirow{2}[2]{*}{5 m} & Simulation & 0.0627 & 0.0072 & 0.0659 \\
          & Fine-tuned & 0.0568 & 0.0029 & 0.0299 \\
    \midrule
    \multirow{2}[2]{*}{7 m} & Simulation & 0.0784 & 0.0131 & 0.1198 \\
          & Fine-tuned & 0.0776 & 0.0113 & 0.1097 \\
    \midrule
    \multirow{2}[2]{*}{10 m} & Simulation & 0.1907 & 0.1337 & 0.7086 \\
          & Fine-tuned & 0.1763 & 0.1052 & 0.6407 \\
    \bottomrule
    \end{tabular}%
    \label{tab:fine_tune_ble}%
\end{table}%

A comparison between Table~\ref{tab:simulation_results_combined} and Table~\ref{tab:exp_results_ble} demonstrates that there is an inevitable simulation-to-reality gap for BLE information hiding. Given that the secret embedder and extractor are trained using the simulation data, fine-tuning model weights with real-collected signals can improve the system performance in practical wireless environments. The effect of sim-to-real fine-tuning is demonstrated in Table~\ref{tab:fine_tune_ble}, where the weights of the secret extractor are fine-tuned using the dataset collected in a short-range LOS setting at a distance of 2~m with an average SNR of 22.54~dB. It can be observed that the $\mathrm{SBER}_{raw}$, $\mathrm{SBER}_{fec}$, and $\mathrm{SMER}$ for other datasets can be significantly reduced after fine-tuning is applied. In the best case, the $\mathrm{SMER}$ reduces from 0.0659 to 0.0299 when evaluated on the dataset collected at a 5~m distance.

\subsubsection{Compatibility with Smartphones}\label{sec:compatibility_smartphone}

\begin{figure}[!t]
    \centering
    \includegraphics[width = 3.4in]{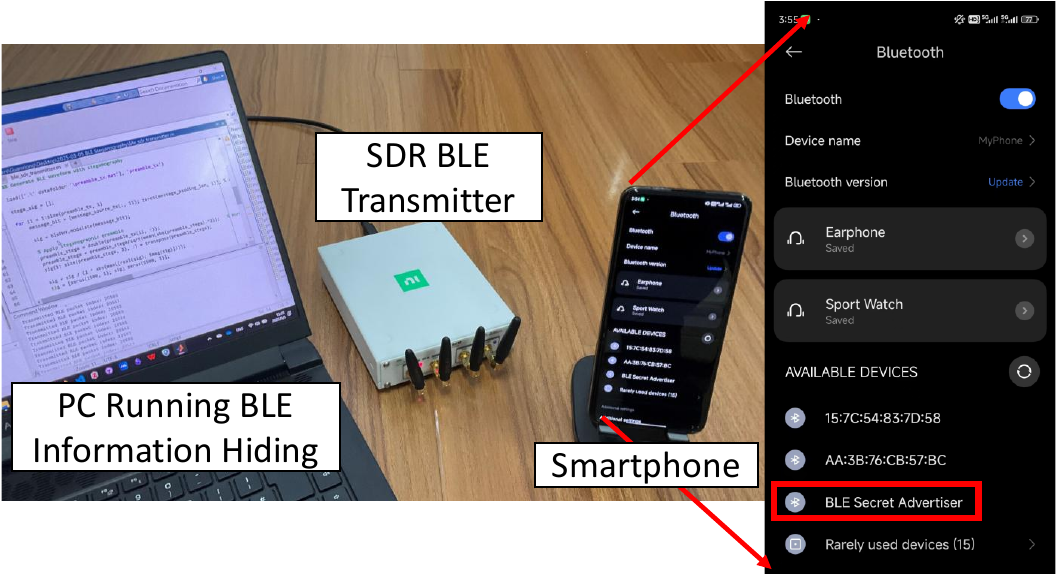}
    \caption{BLE signal reception using a smartphone. The transmitter is an SDR broadcasting BLE advertisements, i.e., random 60-bit secrets are embedded. The receiver is an Android smartphone. The SDR-BLE transmitter, named `BLE Secret Advertiser', is discoverable by the smartphone.}
    \label{fig:phone_reception}
\end{figure}

This subsection provides experimental evidence that the PHY information hiding does not affect the primary communication link, i.e., normal BLE packet reception and decoding. As illustrated in Fig.~\ref{fig:phone_reception}, we use a USRP B210 SDR to broadcast BLE advertisements at channel 37. A random 60-bit secret is embedded within the preamble field of each BLE packet. Concurrently, an Android smartphone is utilized to capture these BLE advertisements. 
It is evident from the phone screen that the BLE packets are successfully received and decoded. The phone screen displays that the SDR-based BLE advertiser is discoverable in instances where 60-bit secrets have been embedded. This proves the claim that the proposed PHY information hiding scheme does not affect the regular packet reception and decoding process. 

In the BLE information hiding system, Alice and Bob can establish a covert link at the PHY layer to transmit secret messages. While others within range can detect the BLE wireless packets using their smartphones, only Bob has the ability to extract the embedded covert secrets. This highlights that the designed PHY information hiding scheme exhibits excellent covertness and stealthiness. 

\subsection{Impact on CFO Estimation Results}\label{sec:impact_cfo_results}

The simulation results in Sections~\ref{sec:impact_cfo_lora} and~\ref{sec:impact_cfo_ble} have demonstrated that the proposed PHY information hiding has negligible impact on CFO estimation. To further validate this conclusion with real hardware, we plot the estimated CFO values from the experimentally collected secret-embedded packets as histograms.

\begin{figure}[!t]
	\centering
	\subfloat[]{\includegraphics[width=1.7in]{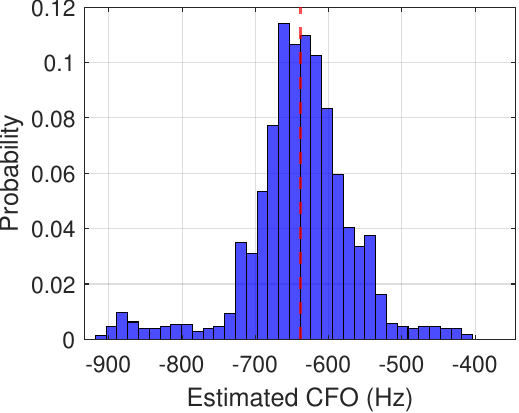}}
	\subfloat[]{\includegraphics[width=1.7in]{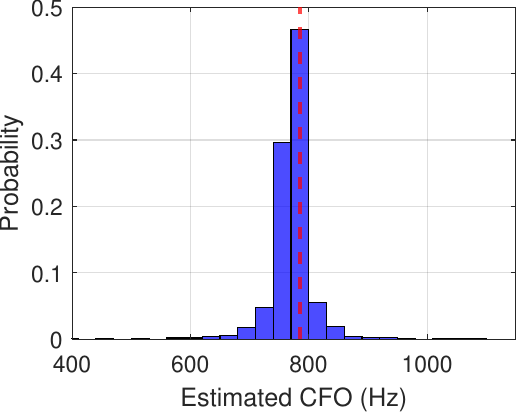}}
	\caption{Distribution of estimated CFO values from experimentally collected secret-embedded packets. The red dashed line denotes the mean CFO of the collected packets. (a) LoRa CFO estimation results. (b) BLE CFO estimation results.}
	\label{fig:cfo_experimental}
\end{figure}

The results are presented in Fig.~\ref{fig:cfo_experimental}, where the red dashed line denotes the mean CFO of the collected packets. For LoRa, the estimated CFO values range from approximately $-900$~Hz to $-400$~Hz. For BLE, the estimated CFO values range from approximately 400~Hz to 1000~Hz. These CFO values primarily originate from the frequency mismatch between the transmitter and receiver crystal oscillators, rather than from the secret embedding algorithm.

Moreover, the estimated CFO magnitudes are negligible relative to the CFO tolerance of the respective protocols. As specified in the Semtech application note~\cite{semtech2023lora_doppler}, the LoRa modem can tolerate frequency offsets of up to $\pm$80~ppm for SF5-SF10, which corresponds to $\pm$34.6~kHz at 433~MHz. The Bluetooth Core Specification~\cite{bluetooth2024core} requires that the transmitter center frequency deviation shall not exceed $\pm$75~kHz. The observed CFOs of $<$1~kHz in both cases are far below these limits and thus will not affect normal packet demodulation. These experimental results confirm that CFO estimation remains fully functional on secret-embedded preambles in practical wireless environments.

\subsection{Model Complexity and Inference Latency}\label{sec:nn_complexity}

We further evaluate the computational overhead introduced by the neural network-based secret embedder and extractor. The benchmark is conducted using PyTorch on an NVIDIA RTX 4060 GPU. For each model, we report the mean and standard deviation over 50 timed runs. The benchmark uses the same BLE and LoRa configurations as those adopted in the real-world experiments described above.

\begin{table}[!t]
  \centering
  \caption{Complexity and inference latency of the secret embedder and extractor. Latency is reported as mean $\pm$ standard deviation over 50 timed runs.}
  \footnotesize
  \setlength{\tabcolsep}{2.5pt}
  \begin{tabular}{ccccc}
    \toprule
    Protocol & Secret Length & Module & Parameters & Latency \\
    \midrule
    \multirow{2}[2]{*}{BLE} & \multirow{2}[2]{*}{60 bit} 
      & Embedder & 3,436,158 & $0.6646 \pm 0.0917$~ms \\
      & & Extractor & 3,328,136 & $0.7626 \pm 0.2087$~ms \\
    \midrule
    \multirow{2}[2]{*}{LoRa} & \multirow{2}[2]{*}{120 bit} 
      & Embedder & 3,651,318 & $0.6419 \pm 0.1727$~ms \\
      & & Extractor & 3,446,576 & $0.7381 \pm 0.1435$~ms \\
    \bottomrule
  \end{tabular}%
  \label{tab:nn_complexity}%
\end{table}%

The results are summarized in Table~\ref{tab:nn_complexity}. The BLE embedder and extractor contain 3.44M and 3.33M parameters, respectively, while the LoRa embedder and extractor contain 3.65M and 3.45M parameters, respectively. The moderate parameter increase in the LoRa models is mainly caused by the longer secret length, since $N$ affects the input channel dimension of the embedder and the final linear layer of the extractor. For both PHYs, the average inference latency of each neural network module is below 0.8~ms, showing that the proposed framework introduces limited neural-network inference overhead on GPU platforms.

For future deployment in more resource-constrained settings, the computational overhead can be substantially reduced through well-established techniques. These include model compression via quantization and pruning, FPGA-based hardware acceleration, and on-chip neural processing units (NPUs) within baseband chips.

\section{Related Work and Discussion}\label{sec:related_work}

\subsection{Related Work}\label{sec:related_studies}

\textit{\textbf{Information Hiding}} has a long history of concealing secret messages within ordinary media and remains widely employed in modern digital watermarking, fingerprinting, and steganography tasks~\cite{zhu2018hidden, luo2020distortion, tancik2020stegastamp, lu2021large, bai2024information, tang2019cnn, zhang2020udh, yang2018rnn, kaptchuk2021meteor, jois2024pulsar}. Traditionally, information hiding relied on manually-defined rules for embedding and extracting hidden information. For instance, the least significant bit (LSB) algorithm embeds secrets by modifying the LSB of pixel values~\cite{fridrich2001detecting}. 
Recently, deep learning-driven approaches have revolutionized information hiding by employing neural networks to automatically learn embedding and extraction algorithms, thereby improving hiding capacity and robustness. The end-to-end deep learning framework for steganography was first introduced in~\cite{zhu2018hidden}, where a pair of neural networks was used: one for embedding secrets into an image and another for extracting them. This encoder-decoder framework rapidly evolved, incorporating advanced techniques such as generative adversarial networks (GANs)~\cite{zhang2019steganogan, yang2019embedding}, invertible neural networks (INNs)~\cite{jing2021hinet, xu2022robust, lu2021large}, and diffusion models~\cite{yu2024cross, yang2024diffstega}. These deep learning innovations have significantly accelerated the development of information hiding.

\textit{\textbf{Covert Communication}} is another closely related topic~\cite{chen2023covert}. The theory-oriented covert communication studies do not embed additional secrets within the PHY. Instead, the covertness is achieved by reducing the likelihood of normal transmissions being detected by an adversary. They are roughly divided into two categories. The first category relies on directional transmission as a means to facilitate covert communication. Through the utilization of multiple antennas~\cite{shmuel2021multi,du2022performance,zheng2019multi}, RIS~\cite{lu2020intelligent, wang2021intelligent,zhou2021intelligent}, or mmWave~\cite{zhang2021joint, wang2021covert, xiao2024star} technologies for beamforming, transmitters can concentrate signals specifically toward the intended legitimate receiver. This technique markedly reduces the probability of interception by eavesdroppers located in other directions, thus improving communication security. Nevertheless, these methods become ineffective when eavesdroppers are geographically positioned near the legitimate receiver. Additionally, controlling side-lobe leakage remains challenging, potentially undermining the effectiveness of such approaches. Another category of studies aims to interfere with the eavesdropper. Specifically, by deploying a friendly jammer to emit artificial noise, the transmitter can interfere with adversaries without affecting legitimate receivers~\cite{zheng2021wireless,zhang2021covert, huang2021jamming}. However, these approaches often require prior knowledge of the eavesdropper, e.g., channel state information, which is difficult to acquire in practical scenarios. Moreover, these schemes are applicable solely for covert communication and cannot be utilized for watermarking or fingerprinting, since no additional secrets are embedded within PHY. Furthermore, the proposed PHY information hiding approach can coexist with other physical-layer security techniques, such as key generation~\cite{huan2025kpgt,zhang2016key,zhang2018channel,huan2024kerra}, to jointly protect wireless systems at the physical layer.

A few system-oriented studies modify existing modulation schemes to create a covert side channel~\cite{hou2022cloaklora, liu2023lophy, schulz2018shadow, huang2019reliable, bonati2021stealte}. For example, the authors in~\cite{d2019hiding} propose a pseudo-noise asymmetrical shift keying (PN-ASK) scheme, which integrates amplitude modulation into PSK techniques. However, these customized modulation-based solutions are only compatible with a specific modulation scheme or wireless protocol, thus exhibiting limited generalizability.

\subsection{Applications of PHY Information Hiding}\label{sec:potential_applications}

The proposed PHY information hiding framework embeds an arbitrary binary secret into the emitted radio waveform.
This waveform-level embedding can support two representative applications.

\textit{\textbf{RF Fingerprint Embedding.}} A representative application is RF fingerprint embedding for PHY authentication.
Secret information is attached to the preamble waveform as a complementary identifier; it does not replace cryptographic authentication.
A static embedded identifier is vulnerable to replay, because an adversary can capture the preamble I/Q samples and retransmit them later.
This limitation can be addressed at the protocol level by a freshness mechanism that makes the embedded secret packet-varying, e.g., timestamp, sequence number, or nonce.

\textit{\textbf{Covert Side Channel.}} Another application is a covert side channel that coexists with the primary communication link.
Because the secrets are concealed in the preamble waveform, commodity receivers that only decode the payload remain unaware of the hidden transmission, while the intended receiver recovers the secrets with the paired secret extractor.
As a related reference, CloakLoRa carries a 30-bit covert payload in each LoRa packet by modulating chirp amplitudes~\cite{hou2022cloaklora}, and the proposed LoRa prototype embeds 120 secret bits in the preamble.

\subsection{Discussion}\label{sec:discussion}
The proposed PHY information hiding is fundamentally different from the related studies in Section~\ref{sec:related_studies}.
Compared to prior theory-oriented studies, it does not impose constraints on the geographic locations of legitimate receivers or potential eavesdroppers~\cite{shmuel2021multi,du2022performance,lu2020intelligent,wang2021intelligent,zhang2021joint,wang2021covert}. Furthermore, it does not require prior channel knowledge about the receiver or eavesdropper~\cite{zheng2021wireless,zhang2021covert,huang2021jamming}, thereby enhancing its feasibility and ease of deployment in real-world wireless communication systems. In contrast to existing system-oriented methods that customize specific modulation schemes or wireless protocols~\cite{hou2022cloaklora,liu2023lophy,schulz2018shadow,huang2019reliable,bonati2021stealte}, we employ a pair of neural networks to learn secret embedding directly within preamble waveforms. This design methodology is protocol-agnostic. Specifically, regardless of the underlying wireless protocol, whether LoRa, Bluetooth, ZigBee, or others, the deep learning framework can automatically learn an effective PHY information hiding strategy directly from simulation data.

However, a limitation of the proposed PHY information hiding scheme is that its performance degrades significantly when the SNR falls below 10~dB, as validated in both simulations and experiments in Sections~\ref{sec:simulation_evaluation} and~\ref{sec:real_world_evaluation}. This behavior is expected, since the secrets are embedded as subtle distortions within the PHY waveform, which become obscured by noise in low-SNR scenarios.

\section{Conclusion}\label{sec:conclusion}

This paper proposes a practical and protocol-agnostic PHY information hiding methodology, which is fundamentally different from prior studies. The design methodology is inspired by the deep learning-based encoder-decoder framework that is widely adopted in other fields, e.g., image information hiding. Specifically, Alice utilizes a neural network to embed secrets within the packet preamble waveforms by introducing subtle and imperceptible distortions. Subsequently, Bob uses another paired neural network extractor to parse the secrets from the packet preamble. We propose a simulation-driven joint training scheme and a sim-to-real fine-tuning technique to enhance the performance when deployed on real hardware transceivers.
We provide SDR hardware prototypes based on LoRa and BLE protocols, demonstrating the excellent generalizability of this scheme. The results validate that the proposed PHY information hiding methodology can establish an undetectable covert channel without affecting the quality of the primary communication links.
For future work, we plan to extend the framework to wideband systems such as Wi-Fi and 5G. This extension is more challenging than the LoRa and BLE settings studied in this paper, because wideband transmissions are more sensitive to frequency-selective fading and multipath-induced distortions. At the same time, Wi-Fi and 5G receivers typically include standard channel estimation and equalization modules. A promising direction is therefore to incorporate these modules to improve the secret decoding performance.




\bibliographystyle{IEEEtran}
\bibliography{IEEEabrv,mybibfile}

\end{document}